\documentclass[reprint,amsmath, superscriptaddress, amssymb, aps, prb]{revtex4-2}
\usepackage{lipsum}  
\usepackage{stackengine}
\usepackage{graphicx}
\usepackage{amsmath}
\usepackage{soul}
\usepackage{dcolumn}
\usepackage{comment}
\usepackage{bm}
\usepackage{color}
\usepackage[normalem]{ulem}
\begin{document}
\renewcommand{\thesection}{\Roman{section}}
\preprint{APS/123-QED}

\title{Microscopic origin of anomalous interlayer exciton transport in van der Waals heterostructures}
\author{Daniel Erkensten}
\affiliation{Department of Physics, Chalmers University of Technology, 41296 Gothenburg, Sweden}
\author{Samuel Brem}
\affiliation{Department of Physics, Philipps-Universit{\"a}t Marburg, 35037 Marburg, Germany}
\author{Ra\"ul Perea-Caus\'in}
\affiliation{Department of Physics, Chalmers University of Technology, 41296 Gothenburg, Sweden}
\author{Ermin Malic}
\affiliation{Department of Physics, Philipps-Universit{\"a}t Marburg, 35037 Marburg, Germany}
\affiliation{Department of Physics, Chalmers University of Technology, 41296 Gothenburg, Sweden}
\begin{abstract}
Van der Waals heterostructures constitute a platform for investigating intriguing many-body quantum phenomena. In particular, transition-metal dichalcogenide (TMD) hetero-bilayers host long-lived interlayer excitons which exhibit permanent out-of-plane dipole moments.  Here, we develop a microscopic theory for interlayer exciton-exciton interactions including both the dipolar nature of interlayer excitons as well as their fermionic substructure, which gives rise to an attractive fermionic exchange. We find that these interactions contribute to a drift force resulting in highly non-linear exciton propagation at elevated densities in the MoSe$_2$-WSe$_2$ heterostructure. We show that the propagation can be tuned by changing the number of hBN spacers between the TMD layers or by adjusting the dielectric environment. In particular, although counter-intuitive, we reveal that interlayer excitons in free-standing samples propagate slower than excitons in hBN-encapsulated TMDs - due to an enhancement of the net Coulomb-drift with stronger environmental screening. Overall, our work contributes to a better microscopic understanding of the interlayer exciton transport in technologically promising atomically thin semiconductors. 
\end{abstract}
\maketitle

\begin{figure}[t!]
    \includegraphics[width=\columnwidth]{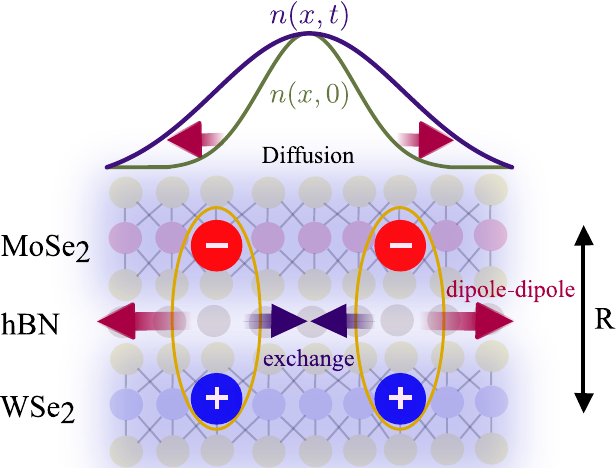}
    \caption{Schematic illustration of the repulsive dipole-dipole interaction (red arrows) and attractive exchange interaction (blue arrows) between interlayer excitons in the exemplary  MoSe$_2$-WSe$_2$ heterostructure. Dipole-dipole repulsion plays a crucial role at elevated densities and leads to a strong drift of the interlayer exciton distribution $n(x,t)$.}
    \label{schematic}
\end{figure}

In recent years, van der Waals heterostructures have emerged as a new class of 2D materials providing a promising playground for studying strong correlations and exotic phases of matter \cite{liu2016van, novoselov20162d, brem2020tunable, brem2022terahertz}. These structures may be formed by stacking transition-metal dichalcogenide (TMD) monolayers on top of each other, which enables the formation of long-lived interlayer excitons \cite{rivera2015observation, ovesen2019interlayer, merkl2019ultrafast, peimyoo2021electrical}, i.e. Coulomb-bound electron-hole pairs where the electronic constituents reside in different layers, cf. Fig. \ref{schematic}. Due to the large separation between electrons and holes, interlayer excitons exhibit permanent out-of-plane dipole moments. These result in strong repulsive inter-excitonic interactions, which are of crucial importance for many-body phenomena in the quantum regime \cite{wang2019evidence, yuan2020twist, bondarev2021crystal, sun2021excitonic}. 

In particular, the effect of strong dipole repulsion has been observed in photoluminescence measurements \cite{nagler2017interlayer, yuan2020twist}, where the interaction becomes manifest in density-dependent blue-shifts of the emission energy with increasing pump power. Moreover, recent experimental studies \cite{yuan2020twist, sun2021excitonic} display the crucial impact of repulsive exciton-exciton interactions in interlayer exciton propagation and transport at elevated electron-hole densities. In this context, highly non-linear exciton diffusion was observed and attributed to the net drift flux of interlayer excitons caused by the repulsive interactions. However, despite the first experiments displaying the importance of the interlayer exciton-exciton interaction in van der Waals heterostructures, its microscopic origin and the role of the exchange coupling and excitonic screening has remained elusive. Furthermore, strategies on how to experimentally control the Coulomb-induced exciton drift have not been systematically studied so far. \\

In this work, we present a microscopic theory of exciton-exciton interactions and study their impact on interlayer exciton propagation in the exemplary MoSe$_2$-WSe$_2$ heterostructure. While most previous studies have focused on dipole-dipole repulsion explaining the non-linear exciton diffusion \cite{sun2021excitonic, nagler2017interlayer, shahnazaryan2021nonlinear}, we show that it is of crucial importance to also consider the fermionic exchange interactions \cite{schindler2008analysis, tassone1999exciton, rivera2016valley} and excitonic screening \cite{ropke1979influence, steinhoff2017exciton}, cf. Fig. \ref{schematic}. We reveal that the interplay of these partially counteracting mechanisms determines the overall spatiotemporal dynamics of interlayer excitons. In particular, we find  that the repulsive interlayer exciton-exciton interaction is considerably reduced by the attractive exchange interaction. These interactions generate a net drift force which leads to an increase in the effective diffusion coefficient by up to an order of magnitude at high exciton densities, resulting in a highly non-linear interlayer exciton propagation. Moreover, we find that the latter can be tuned by changing the interlayer distance, e.g. through the inclusion of hBN spacers (Fig. \ref{schematic}), thereby enhancing the repulsive dipolar interactions between interlayer excitons. Finally, we predict a counter-intuitive dependence on the dielectric environment, where free-standing heterostructures  exhibit a smaller exciton drift, despite the stronger Coulomb interaction between individual charges in a purely classical picture. 

\section*{Theoretical model}
To describe intra- and interlayer exciton-exciton interactions in a microscopic and material-specific way we combine the exciton density matrix formalism \cite{brem2020phonon, erkensten2021exciton, katsch2018theory} with  density functional theory calculations \cite{kormanyos2015k}. First, we  derive the mean-field Hamilton operator $H=H_0+H_{x-x}$, consisting of the free part ($H_0$) and the inter-excitonic interaction part ($H_{x-x}$). We express the Hamiltonian in an excitonic basis with $ H_0= \sum_{\alpha, \bm{Q}}E^{\alpha}_{\bm{Q}}X^{\dagger}_{\alpha, \bm{Q}}X_{\alpha, \bm{Q}}$ and 
\begin{equation}
H_{x-x}= \sum_{\substack{\alpha, \beta \\ \bm{Q}, \bm{q}}}W^{\alpha\beta}_{\mathrm{mf}, \bm{q}, \bm{Q}}X^{\dagger}_{\alpha, \bm{Q}-\bm{q}}X_{\beta, \bm{Q}} \ .
\label{xxhamilton}
\end{equation}
 Here,  $X^{(\dagger)}_{\alpha, \bm{Q}}$ are excitonic operators creating or annihilating an exciton in the state $\alpha=1s,2p,2s...$ with the center-of-mass momentum $\bm{Q}$. The exciton dispersion $E^{\alpha}_{\bm{Q}}=\frac{\hbar^2\bm{Q}^2}{2M}+E^{\alpha}$ includes the exciton binding energy $E^{\alpha}$, which is obtained along with the associated excitonic wave functions $\varphi_{\alpha, \bm{q}}$ by solving the Wannier equation \cite{kira2006many}. The appearing exciton mass $M$  is obtained from DFT calculations \cite{kormanyos2015k}. The full derivation of the mean-field Hamiltonian and the used DFT parameters can be found in the Supplementary Material.

The interaction term, $H_{x-x}$, contains the mean-field exciton-exciton matrix element, $W^{\alpha\beta}_{\mathrm{mf}, \bm{q}, \bm{Q}}$ which includes direct and exchange contributions. In general, both  contributions are momentum-dependent, but the matrix element for 1s excitons is approximately constant in the long-wavelength limit and reads 
\begin{equation}
\lim_{\bm{q}, \bm{Q}\rightarrow 0} W^{1s-1s}_{\mathrm{mf}, \bm{q}, \bm{Q}}=n_x(g_{d-d}+g_{x-x}) \ .
\end{equation}
The full expression for the momentum-dependent matrix element is given in the Supplementary Material. Note that the full momentum-dependent matrix elements have been investigated in detail in a previous study \cite{erkensten2021exciton}. In this work we consider low temperatures and cold exciton distributions peaked around $\bm{Q}=0$, and therefore restrict our studies to the long-wavelength approximation of the exciton-exciton interaction. Moreover, we consider interactions between the mostly occupied 1s exciton states, neglecting the coupling to higher-order exciton states. 
We have introduced the exciton density $n_x$ and the dipole-dipole interaction strength $g_{d-d}$, which reads 
\begin{equation}
    g^{(X)}_{d-d}=0, \ \ \ \  g^{(IX)}_{d-d}=\frac{e^2}{2\epsilon_0}\left(\frac{d_1}{\epsilon^{(1)}_{\perp}}+\frac{d_2}{\epsilon^{(2)}_{\perp}}+\frac{2R}{\epsilon_R}\right) \ 
    \label{eqdd}
\end{equation}
for intralayer ($X$) and interlayer ($IX$) excitons.
Here, $e$ is the electron charge, $d_{i}$ the TMD layer thickness, $\epsilon^{(i)}_{\perp}$ the out-of-plane component of the dielectric tensor \cite{laturia2018dielectric} of the TMD layer $i=1,2$, $R$ the layer separation and $\epsilon_R$ the effective dielectric constant for a spacer with the thickness $R$, cf. Fig. \ref{schematic}. While the dipole-dipole interaction between ground state intralayer excitons vanishes, for interlayer excitons we find a considerable repulsive ($>0$) interaction that can be interpreted as a classical dipole-dipole coupling \cite{de2001exciton, erkensten2021exciton}. 

Since excitons are composite bosons consisting of electrons and holes, exchange of fermionic constituents \cite{tassone1999exciton, katsch2018theory} also has to be included when considering exciton-exciton interactions. In contrast to the classical dipole-dipole coupling, the exchange interaction strength depends strongly on the excitonic wave functions and reads 
\begin{gather}
\begin{aligned}
    g_{x-x}=A\sum_{\substack{\bm{k}, \bm{k'}}}|\varphi_{ \bm{k'}}|^2\bigg(2V^{eh}_{\bm{k}}\varphi^*_{\bm{k}+\bm{k'}}\varphi_{\bm{k'}} -\sum_{\lambda=e,h} V^{\lambda\lambda}_{\bm{k}}|\varphi_{\bm{k'}+\bm{k}}|^2\bigg) , 
    \label{exchange}
\end{aligned}
\end{gather}
where $A$ is the crystal area. Note that the appearing electronic Coulomb matrix elements $V^{\lambda\lambda}_{\bm{k}}$ and $V^{\lambda\bar{\lambda}}_{\bm{k}}$ ($\lambda\neq\bar{\lambda}$) are proportional to $1/A$, such that $g_{x-x}$ is independent of $A$. The first term in Eq. \eqref{exchange} is proportional to the electron-hole interaction ($V^{eh}_{\bm{k}}$), while the second is determined by the electron-electron ($V^{ee}_{\bm{k}}$) and hole-hole ($V^{hh}_{\bm{k}}$) interactions \cite{tassone1999exciton, bobrysheva1972bi}. Importantly, the contributions $V^{eh}$ and $V^{ee}/V^{hh}$ come with different signs, reflecting the attractive and repulsive nature of the electron-hole and  electron-electron (hole-hole) coupling, respectively. 

The nature of the exchange part of the exciton-exciton interaction between intralayer  and interlayer excitons is fundamentally different. In particular, it holds that $V^{eh}_{\bm{k}}\approx{V^{hh}_{\bm{k}}}\approx{V^{ee}_{\bm{k}}}$ ($V^{eh}_{\bm{k}}<<V^{ee}_{\bm{k}}, V^{hh}_{\bm{k}}$) for interactions between intralayer (interlayer) excitons, i.e. the electron-hole interaction is significantly weaker than the Coulomb repulsion between individual electrons or holes when electrons and holes are spatially separated in different layers. As a consequence, the exchange interaction becomes more attractive (negative) as the vertical separation of electrons and holes is increased \cite{PhysRevB.86.115324}. In contrast, the exchange interaction between intralayer excitons is always repulsive (positive) \cite{ciuti1998role}.

When evaluating the electronic Coulomb matrix elements ($V^{eh}_{\bm{k}}, V^{ee}_{\bm{k}}, V^{hh}_{\bm{k}}$) we take into account: \textbf{i)} the finite thickness of the TMD sample and the dielectric screening due to the surrounding substrate and the TMD monolayers themselves, and \textbf{ii)} the excitonic screening present at elevated electron-hole densities. The first (background) contribution is obtained as a generalized Rytova-Keldysh potential \cite{ovesen2019interlayer, keldysh1979coulomb, rytova1967screened} and the latter is calculated using static excitonic polarizabilities via an excitonic Lindhard model \cite{steinhoff2017exciton, ropke1979influence}, inducing a density-dependence in the exciton-exciton interaction (see Supplementary Material for more details). Note, however, that the exciton wave functions are still taken as density-independent, which is assumed to hold as long as the exciton-exciton interaction energy is smaller than the binding energy and can be seen as a perturbative correction to the exciton energy. This assumption is justified in this work as we consider densities at which most electron-hole pairs are bound as excitons, i.e. densities below the exciton Mott transition at $\sim10^{13}$ $\mathrm{cm}^{-2}$ \cite{steinhoff2017exciton}. The negligence of free carrier screening allows us to obtain a theory expressed entirely in excitonic quantities. Note that although excitons (being effectively neutral quasi-particles) are only weakly polarizable compared to the free plasma, we find that the inclusion of higher-lying p-states, e.g. 1s-2p exciton transitions, gives rise to a significant screening stemming from bound states \cite{steinhoff2017exciton} (cf. the Supplementary Material). Finally, we point out that the mean-field treatment of excitons introduced here, considering excitons as pure and independent bosons with the fermionic exchange effects captured via the excitonic Coulomb matrix elements, neglects the formation of biexcitons or higher order particle clusters. Moreover, the theory is expected to hold up to first order in exciton density, $n_x$. In particular, this translates to fulfilling the condition $n_x a_B^2<<1$, corresponding to exciton densities $n_x<<10^{14}$ $\mathrm{cm}^{-2}$ assuming an exciton Bohr radius $a_B\sim 1$ nm \cite{de2001exciton}. 

\section*{Density-dependent spectral shifts}

We now make use of the Hamiltonian in \eqref{xxhamilton} to determine the density-dependent spectral shifts of exciton resonances, which are accessible e.g. in photoluminescence spectra \cite{nagler2017interlayer, yuan2020twist}. These shifts can be microscopically calculated directly from the equation of motion for the bright exciton polarization, $P_{1s}=\langle X^{\dagger}_{1s, \bm{Q}=0}\rangle$, which on a Hartree-Fock level reads (cf. the Supplementary Material): 
\begin{equation}
    \dot{P}_{1s}=\frac{i}{\hbar}\bigg(E^{1s}+\Delta E(n_x)\bigg)P_{1s} 
\end{equation}
with $n_x=\frac{1}{A}\sum_{\bm{Q}}\langle X^{\dagger}_{\bm{Q}}X_{\bm{Q}}\rangle$.  
Here, we introduced the density-dependent energy renormalization
\begin{equation}
\Delta E (n_x)\equiv (g_{d-d}+g_{x-x})n_x+\Sigma_{CH} \ . 
\label{energyren}
\end{equation}
The last term describes the  Coulomb-hole contribution \cite{hedin1965new}, which explicitly takes into account density-dependent screening effects of the electronic band gap. It reads $\Sigma_{CH}=\sum_{\bm{q}} (V^{hh}_{\bm{q}}-\bar{V}^{hh}_{\bm{q}})$, where $\bar{V}^{hh}_{\bm{q}}$ denotes the \emph{unscreened} Coulomb potential with respect to exciton screening.  Hence, within our formalism, the Coulomb-hole term describes the reduction of the exciton energy due to a screening-induced change in the Coulomb renormalization of the filled valence band (cf. the Supplementary Material for details).

\begin{figure}[t!]
    \centering
    \includegraphics[width=\columnwidth]{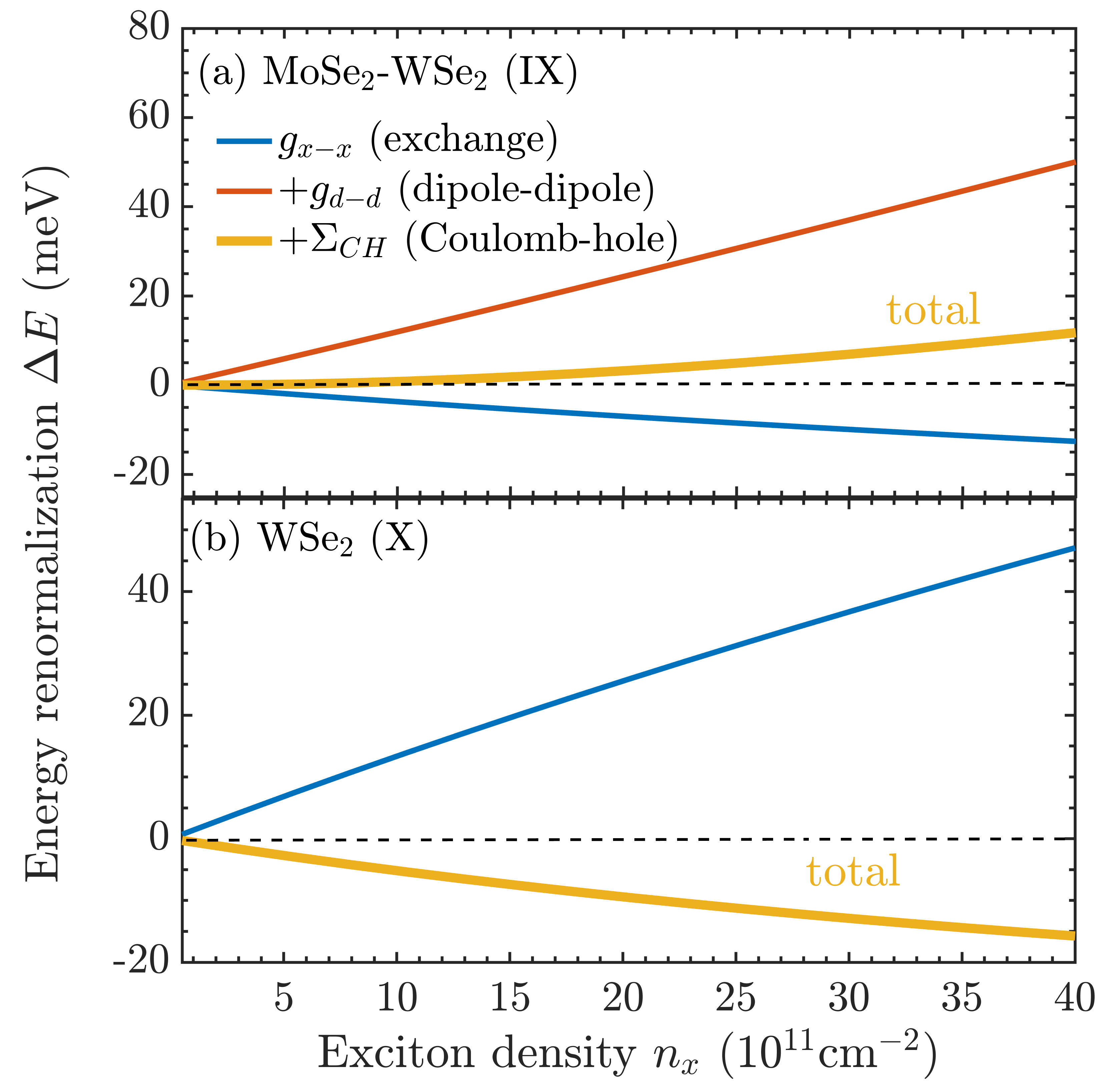}
    \caption{Density-dependent energy renormalization $\Delta E$ due to the exchange interaction ($g_{x-x}$, blue), the dipole-dipole interaction ($g_{d-d}$, red) and the screening-induced Coulomb-hole contribution (yellow). Note that the individual contributions are included additively. \textbf{(a)} Interlayer energy renormalization in the MoSe$_2$-WSe$_2$ heterostructure, resulting in a net blue-shift of the exciton resonance. \textbf{(b)} Intralayer energy renormalization in the WSe$_2$ monolayer, resulting in a net red-shift.}
    \label{lineshift}
\end{figure}

In Fig. \ref{lineshift} we present the corresponding density-dependent energy renormalization for interlayer  (Fig.\ref{lineshift}(a)) and intralayer excitons (Fig.\ref{lineshift}(b)) in the exemplary MoSe$_2$-WSe$_2$ heterobilayer and WSe$_2$ monolayer, respectively. Due to the strong dipole-dipole interaction ($g_{d-d}$) in the heterostructure we obtain a net blue-shift---in agreement with experimental studies \cite{sun2021excitonic, li2020dipolar, nagler2017interlayer}. In contrast, a small red-shift is found in the intralayer case due to the absence of permanent dipole moments. Here, we note that the intralayer exciton-exciton interaction is dominated by quantum-mechanical exchange interactions ($g_{x-x}$), as has been confirmed previously for TMD monolayers \cite{erkensten2021exciton, PhysRevB.96.115409} and quantum wells \cite{ciuti1998role, schindler2008analysis, de2001exciton}. Moreover, as discussed above, the exchange interaction is repulsive (attractive) in the intralayer (interlayer) case. The nature of this interaction  depends strongly on the interplay between the electron-hole and electron-electron/hole-hole Coulomb matrix element, cf. Eq. \eqref{exchange}. 
By including the screening between excitons at elevated densities, the blue-shift of the interlayer exciton resonance is reduced, resulting in a net shift on the order of 15 meV for the considered density range ($0\leq n_x<5\cdot 10^{12}$ $\mathrm{cm}^{-2}$).

Finally, we also observe the predominant role of the Coulomb-hole term in monolayers, compensating for the exchange-induced shift to higher energies and even leading to a net red-shift of intralayer exciton resonances---in agreement with experimental observations \cite{yuan2020twist} and previous microscopic calculations \cite{PhysRevB.98.035434}. Note that correlation effects and free carrier screening are neglected here, which could reduce the shift of exciton resonances further.

In the following, we will focus on spatially inhomogeneous systems, assuming density gradients on much larger scales than the exciton Bohr radius/thermal wave length. In real space, the energy shifts discussed above translate into a spatially varying potential $ \Delta E(n(x,y))$ which leads to a drift of excitons. Thus, we now consider the propagation of interlayer excitons and focus in particular on the impact of exciton-exciton interactions.

\section*{Coulomb-driven exciton propagation}

\begin{figure*}[t!]
    \includegraphics[scale=0.45]{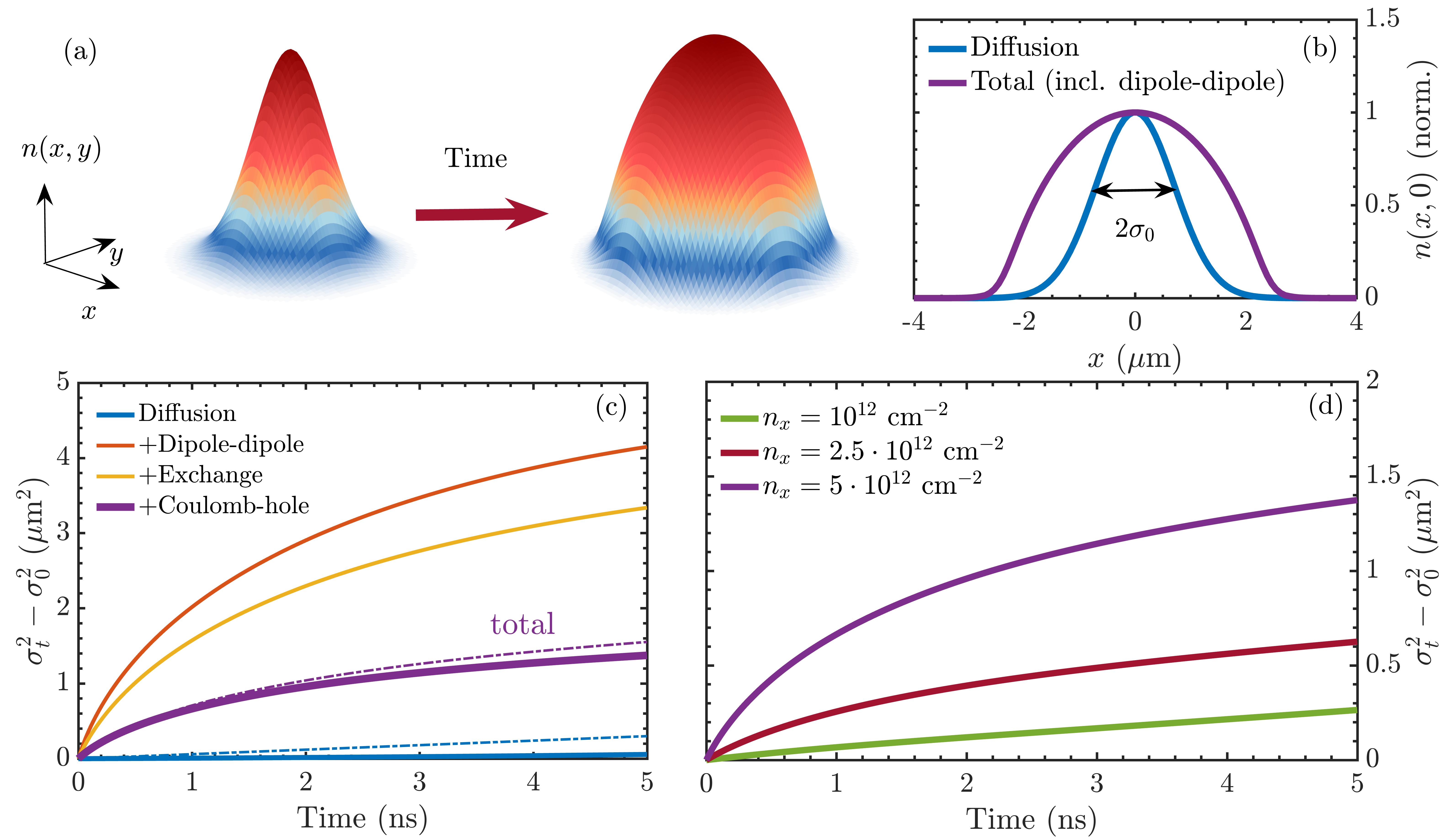}
    \caption{Time evolution of interlayer exciton density $n(x,y,t)$ in the MoSe$_2$-hBN-WSe$_2$ heterostructure at 4.6 K. \textbf{(a)} Exciton distribution initialized as a Gaussian distribution  with an initial variance set to $\sigma_0^2=1$ $\mu\mathrm{m}^2$ and an exciton density of $n_x=5\cdot 10^{12}$ $\mathrm{cm}^{-2}$. As time progresses a significant broadening of the distribution is observed due to the strong repulsive exciton-exciton interaction. \textbf{(b)} Cut of the exciton distribution $n(x,0)$ at $t=5$ ns with (purple) and without (blue) the Coulomb-induced drift. 
    \textbf{(c)} Time-dependent variances $\sigma^2_t$ at the exciton density $n_x=5\cdot 10^{12} \ \mathrm{cm}^{-2}$ taking into account just conventional diffusion (blue) and including Coulomb drift  stemming from the exchange interaction, the dipole-dipole repulsion, and the Coulomb-hole term. The dashed lines show the solution of the drift-diffusion equation assuming a Boltzmann distribution for the interlayer exciton including only diffusion (blue) and including both diffusion and Coulomb-induced drift terms (purple). \textbf{(d)} Impact of the exciton density on the interlayer exciton transport. Anomalous diffusion is observed at densities $n_x> 10^{12}$ $\mathrm{cm}^{-2}$.}
    \label{timeandspace}
\end{figure*}

The spatiotemporal dynamics of excitons can be accessed through the temporal evolution of the exciton Wigner function \cite{perea2019exciton,rosati2021dark, rosati2020negative}, which is directly extracted from the Heisenberg equation of motion for the off-diagonal density matrix $\langle X^{\dagger}_{\bm{Q}} X_{\bm{Q'}}\rangle$. Extending the approach introduced by Hess and Kuhn \cite{hess1996maxwell, hess1996spatio} to excitons, we can quantitatively describe the spatiotemporal evolution of the interlayer exciton density $n(\bm{r},t)$ through the following drift-diffusion equation: 
\begin{multline}
\begin{aligned}
   \dot{n}(\bm{r}, t)&=\nabla\cdot ( D(n(\bm{r},t)) \nabla n(\bm{r},t) )\\
   &+ \mu_m \nabla\cdot( \Delta E(n(\bm{r},t)) \nabla n(\bm{r},t))-\frac{n(\bm{r},t)}{\tau} \ , 
    \label{dde}
\end{aligned}
\end{multline}
with $D(n)=D_0 \frac{T_d}{T}[\mathrm{exp}(T_d/T)-1]^{-1}$, where $D_0$ is the low-density diffusion coefficient, $T_d=\frac{2\pi n  \hbar^2}{M k_B}$  the degeneracy temperature, $\mu_m=\frac{D_0}{k_B T}$ the exciton mobility and $\tau$ the exciton life time  \cite{ivanov2002quantum}. At the cryogenic temperatures ($T<<T_d$) considered in this work, the diffusion coefficient $D$ acquires a strong density-dependence as a result of boson bunching. Here, we have included the  spatially varying exciton-exciton interaction energy $\Delta E(n(\bm{r},t))$ (Eq. \eqref{energyren}) consisting of contributions from the dipole-dipole interaction ($g_{d-d}$), the exchange interactions ($g_{x-x}$) and the Coulomb-hole term ($\Sigma_{CH}$). Note that we disregard exciton-exciton annihilation processes \cite{yu2016fundamental, erkensten2021dark} in this work. These are assumed to be negligible for interlayer excitons, as has been confirmed by recent time-dependent photoluminescence measurements in MoSe$_2$-WSe$_2$ \cite{sun2021excitonic} as well as WS$_2$-WSe$_2$ hetero-bilayers \cite{yuan2020twist}. Additional details on the derivation of \eqref{dde} are found in the Supplementary Material. 

The first term in Eq. \eqref{dde} accounts for the diffusive propagation of excitons and leads to conventional diffusion at low excitation densities with a time-independent propagation speed given by the low-density diffusion coefficient $D_0$according to Fick's law \cite{kulig2018exciton, perea2019exciton}. The second term, which arises due to net repulsive exciton-exciton interactions, leads to a drift flux of interlayer excitons and is therefore denoted as the Coulomb-induced drift. Finally, the third term in \eqref{dde} describes the population decay of the interlayer exciton density. By numerically solving the  drift-diffusion equation (Eq. \eqref{dde}) for an exemplary hBN-encapsulated MoSe$_2$-hBN-WSe$_2$ heterostructure (Fig. \ref{schematic}), we gain microscopic access to the spatiotemporal dynamics of interlayer excitons. In the considered AA-stacked heterostructure we note that the bright KK interlayer exciton state is by far the energetically lowest state due to the large type-II band alignment \cite{PhysRevB.97.165306}, and therefore it is justified to restrict our analysis to this state. Moreover, we can neglect the effects of hybridisation due to the weak interlayer tunneling strength at the K-point \cite{PhysRevB.95.115429, hagel2021exciton}. The inclusion of an hBN spacer with the thickness $R=0.3$ nm additionally allows us to disregard the possibility of interlayer excitons being trapped by the moir{\'e} potential, which is known to affect exciton transport in vdW heterostructures \cite{choi2020moire, PhysRevLett.126.106804,brem2020tunable}. 

Furthermore, we adapt our study to the experimental conditions of Ref. \cite{sun2021excitonic} by setting the temperature to $T=4.6$ K, the low-density diffusion coefficient $D_0=0.15$ $\mathrm{cm}^2/\mathrm{s}$ and the exciton life time $\tau=3.5$ ns obtained from transport measurements. 
In Fig. \ref{timeandspace}(a) we illustrate the exciton density $n(x,y,t)$ for different times $t$. We initialize the exciton distribution as a Gaussian, $n(x,y,0)=n_x \mathrm{exp}(-(x^2+y^2)/\sigma_0^2)$ and specify the initial density $n_x=5\cdot 10^{12}$ $\mathrm{cm}^{-2}$ and laser spot size $\sigma_0^2=1$ $\mu\mathrm{m}^2$. As time progresses, the density develops into a super-Gaussian distribution whose spatial width, given by $\sigma_t^2=\int \bm{r}^2n(\bm{r},t)d\bm{r}/\int  n(\bm{r},t)d\bm{r} $, evolves highly non-linearly with time---a hallmark of \emph{anomalous} diffusion, cf. Fig. \ref{timeandspace}(c). 

Including just the diffusive part of the exciton transport ($\propto D$, cf. Eq. \eqref{dde}), the exciton distribution retains its Gaussian shape, cf. Fig. \ref{timeandspace}(b). However, as the diffusion coefficient decreases rapidly with density at low temperatures (due to boson bunching), the width of the exciton distribution varies sublinearly with time (blue, solid lines in Fig. \ref{timeandspace} (c)). It stays approximately constant when comparing with the case of a constant diffusion coefficient $D_0$, in which the width increases linearly with time, i.e. $\sigma_t^2-\sigma_0^2=4D_0t$ according to Fick's law (blue, dashed lines in Fig. \ref{timeandspace} (c)). Instead, the transport properties of interlayer excitons are governed by exciton-exciton interactions at elevated densities. In particular, the repulsive dipole-dipole interaction leads to the drift of interlayer excitons giving rise to a fast propagation, which is much enhanced relative to diffusion. In particular, we obtain an effective diffusion coefficient of $D_{\mathrm{eff}}=1.5$ $\mathrm{cm}^2/\mathrm{s}$ from the slope of the variance (at $t$=1.5 ns), which is one order of magnitude larger than the low density diffusion coefficient of $D_0=0.15$ $\mathrm{cm}^2/\mathrm{s}$.  
The net impact of the interaction is reduced when also taking into account the exchange interaction---reflecting its attractive nature (cf. Eq. \eqref{exchange})---and the Coulomb-hole contribution (cf. Eq. \eqref{energyren}). The latter significantly reduces the interaction potential at elevated densities, as observed in Fig. \ref{timeandspace}(c), and weakens the Coulomb-induced drift due to dipole-dipole interaction, cf. solid purple lines. For comparison, we provide the time-dependent variance as obtained from solving the drift-diffusion equation assuming a constant diffusion coefficient $D_0$, cf. the dashed purple lines. This corresponds to assuming a Boltzmann distribution for excitons, retrieved by taking the classical limit $T>>T_d$ in Eq. \eqref{dde}. 
As exciton drift completely dominates over diffusion at elevated densities and the exciton mobility is only dependent on the low-density diffusion coefficient, it follows that the solid and dashed lines qualitatively coincide and that the anomalous character of the transport is only weakly dependent on the exciton distribution.

As a consequence of the interplay between the repulsive (dipole-dipole) and attractive (exchange and Coulomb-hole) contributions to the Coulomb-induced drift, considerably high densities are needed to actually observe an anomalous exciton diffusion. The density dependence of the interlayer exciton transport is investigated further in Fig. \ref{timeandspace}(d), where we find that densities larger than $n_x=10^{12}$ $\mathrm{cm}^{-2}$ are required to observe non-linear exciton propagation. At densities $n_x\leq 10^{12}$ $\mathrm{cm}^{-2}$ the propagation becomes purely diffusive and exciton-exciton interactions play a minor role. In particular, at densities $n_x=10^{12}$ $\mathrm{cm}^{-2}$ (green curve) and below, the time-dependent variance $\sigma_t^2-\sigma_0^2$ approaches $4D_0t$ with the low-density diffusion coefficient $D_0$, as expected from Fick's law. 

\begin{figure}[t!]
    \centering
    \includegraphics[width=\columnwidth]{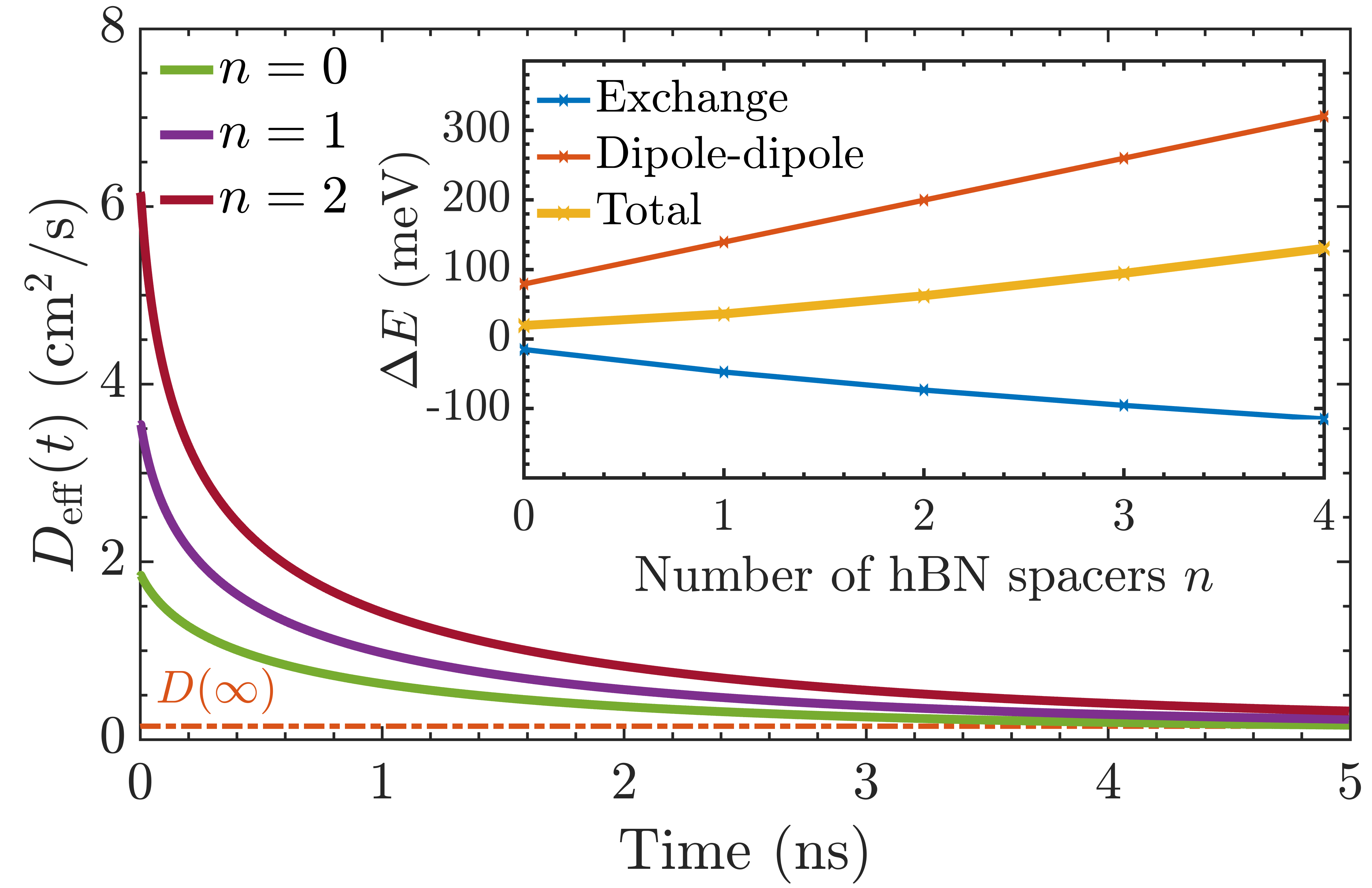}
    \caption{Time-dependent effective diffusion coefficient $D_{\mathrm{eff}}(t)\equiv \frac{1}{4}\frac{d}{dt} \sigma^2_t$ for different numbers of hBN-spacers placed between the two layers in the MoSe$_2$-WSe$_2$ heterostructure for a fixed exciton density $n_x=5\cdot 10^{12}$ $\mathrm{cm}^{-2}$. A drastic increase in the diffusion coefficient at initial times is found with the number of hBN-spacers. At large times ($t\rightarrow\infty$) the coefficients approach the indicated low-density coefficient $D=0.15$ $\mathrm{cm}^2/\mathrm{s}$ (dashed orange line). The inset illustrates the dependence of the exciton-exciton interaction energy $\Delta E$ (cf. Eq. \eqref{energyren}) on the layer separation and shows that the total interaction (yellow line)  is boosted with an increasing number of hBN spacers. }
    \label{hbnspacer}
\end{figure}

\section*{Tunability of the Coulomb-induced interlayer exciton drift}
Having determined the microscopic nature of exciton-exciton interactions in van der Waals heterostructures and their impact on exciton propagation, we now investigate how the propagation can be tuned. The possibility to control the exciton transport is of technological relevance for the design of devices based on atomically thin semiconductors.  In particular, successful control of exciton transport paves the way for creating excitonic devices, such as transistors \cite{unuchek2018room}. In the following, we identify and investigate three experimentally accessible knobs to tune the propagation of excitons: \textbf{i)} interlayer distance, \textbf{ii)} dielectric environment, and \textbf{iii)} excitation spot size.\\

\textbf{Interlayer distance:}
The dipole-dipole interaction between interlayer excitons is highly tunable with the layer separation, $R$ (cf. Fig. \ref{schematic}), as observed in recent experimental studies \cite{yuan2020twist, sun2021excitonic}. In particular, we find that the dipole moment of interlayer excitons is enhanced by more than 50 $\%$ by including an hBN spacer into the heterostructure. As a direct consequence, the dipole-dipole interaction (Eq. \eqref{eqdd}) can be boosted by increasing the layer separation. Here, we show how a change of the interlayer separation affects the Coulomb-induced drift on microscopic footing, including the impact on the exchange interaction as well as exciton screening.

In Fig. \ref{hbnspacer}, we show the time-dependent effective diffusion coefficients $D_{\mathrm{eff}}(t)\equiv \frac{1}{4}\frac{d}{dt} \sigma^2_t$ for an increasing number of hBN spacers. We find that interlayer excitons propagate faster when the interlayer separation is larger. This reflects the increased contribution of the  dipole-dipole repulsion, as shown in the inset in Fig. \ref{hbnspacer}. Note that the exchange interaction is also very sensitive to changes in the interlayer distance---as it crucially depends on the electron-hole interaction---and it partially counteracts the classical dipole-dipole repulsion (cf. inset). 
In particular, we find that the reduction of the electron-hole interaction with interlayer distance results in a decrease in the exciton binding energy from $\approx{90}$ meV to $\approx{78}$ meV when comparing hBN-encapsulated MoSe$_2$-WSe$_2$ heterostructures with and without a hBN spacer respectively, indicating that the exciton stays tightly bound in the presence of a small number of spacers.
Importantly, the Coulomb-induced drift of excitons gives rise to a significant enhancement of the effective diffusion coefficient at initial times, when the exciton density is at its largest. The diffusion coefficients range from $D_{\mathrm{eff}}\approx{6}$ $\mathrm{cm}^{2}/\mathrm{s}$ to $D_{\mathrm{eff}}\approx{2}$ $\mathrm{cm}^{2}/\mathrm{s}$ when considering heterostructures with and without hBN spacers, respectively. This corresponds to an enhancement of more than an order of magnitude compared to the low-density diffusion coefficient of $D=0.15$ $\mathrm{cm}^2/\mathrm{s}$ (indicated by the dashed line in Fig. \ref{hbnspacer}). At larger times, the exciton density drops as excitons have propagated away from the excitation spot, leading to the low density regime where the impact of exciton-exciton interaction is small and conventional diffusion is predominant.   \\

\begin{figure}[t!]
    \centering
    \includegraphics[width=\columnwidth]{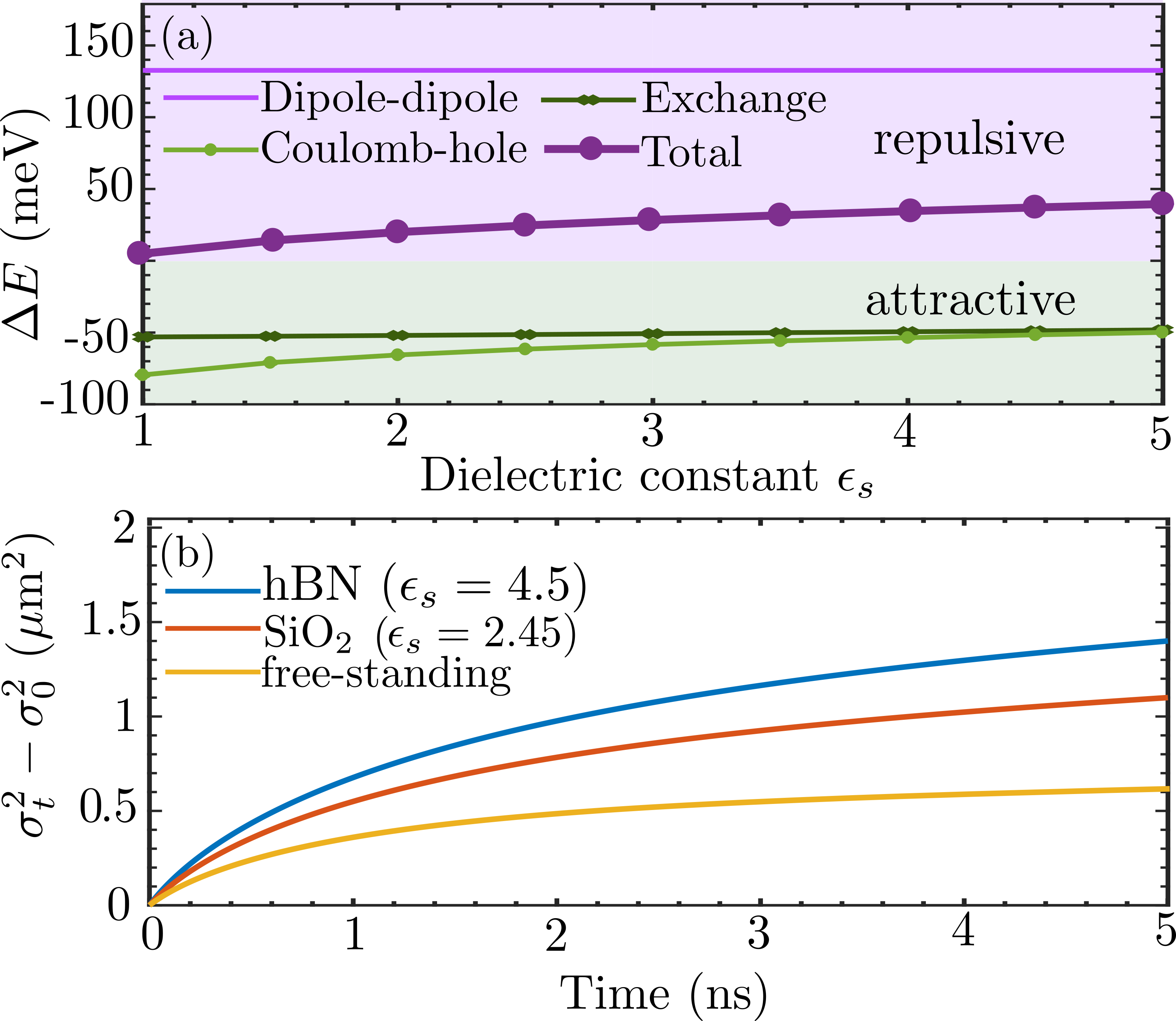}
    \caption{Impact of dielectric environment on spectral shift and transport of interlayer excitons at a fixed exciton density $n_x=5\cdot 10^{12}$ $\mathrm{cm}^{-2}$. \textbf{(a)} Screening dependence of the energy renormalization $\Delta E$ distinguishing different contributions. The total exciton-exciton interaction (purple line) shows only a weak dependence that can be traced back to the Coulomb-hole contribution (bright green line).
    \textbf{(b)} Temporal evolution of the variance $\sigma_t^2-\sigma_0^2$ as a function of time for different surrounding substrates.
    }
    \label{xxb}
\end{figure}

\textbf{Dielectric engineering:} 
Besides varying the number of hBN spacers between the TMD monolayers, the exciton-exciton interaction can be tuned by changing the surrounding environment, i.e. by modifying the dielectric constant of the surrounding material, $\epsilon_s$. We find that the dipole repulsion and the exchange coupling exhibit only a weak dependence on the dielectric environment. Figure \ref{xxb}(a) illustrates the changes in the energy renormalization $\Delta E$ (Eq. \eqref{energyren}) as a function of the dielectric constant. In the considered long-wavelength limit, the dipole-dipole interaction is independent of screening (cf. Eq. \eqref{eqdd}) and the exchange interaction (cf. Eq. \eqref{exchange}) leads to a very small change in $\Delta E$ (in the range of just a few meV from the free-standing to the hBN-encapsulated samples). This is in stark contrast to the conventional Coulomb interaction between electrons and holes, in which the Coulomb matrix elements are approximately inversely proportional to $\epsilon_s$. In general, the exciton-exciton interaction is determined by the Coulomb potentials $V^{eh}$, $V^{ee}$ and $V^{hh}$, but it additionally depends on form factors that scale with the excitonic wave functions in momentum space $\sim|\varphi_{\bm{k}}|^4$ (cf. Eq. \eqref{exchange}). As the dielectric constant of the environment is increased, the Coulomb interaction is weakened, which makes the excitonic wave function less localized in real space. This means, however, that the wavefunction becomes narrower in momentum space resulting in larger form factors. 
Visually speaking, when the Bohr radius of the exciton is increased it is more likely that two excitons occupy the same space, which increases the interaction strength.
The competition between increased exciton wave function overlap and weakened Coulomb interactions is the microscopic origin of the observed weak screening dependence of $g_{d-d}$ and $g_{x-x}$, cf. Fig. \ref{xxb}(a). 

While the dielectric properties of the environment do not significantly impact the strength of the dipole-dipole and exchange interactions, we find that the non-linear time-dependence of the variance $\sigma_t^2$ can still be strongly tuned with dielectric engineering. As we show in Fig. \ref{xxb}(b), hBN-encapsulated samples are seen to result in more anomalous (faster) diffusion than free-standing TMDs. This is a  counter-intuitive result, as one would expect the Coulomb-induced drift to be more efficient in the free-standing case, where the screening of the Coulomb interaction is weak. We can trace back this striking behaviour to the screening-dependence of the Coulomb-hole term (Eq. \eqref{energyren}), cf. the bright green line in Fig. \ref{xxb}(a). In particular, we find that this term can be expressed through the exciton polarizability $\Pi_{\bm{q}}$ according to $\Sigma_{CH}=-\sum_{\bm{q}}\frac{(V^{hh}_{\bm{q}})^2 |\Pi_{\bm{q}}|}{1+V^{hh}_{\bm{q}}|\Pi_{\bm{q}}|}$, where $\Pi_{\bm{q}}$ approximately scales with the ratio of the in-plane exciton transition dipole moment and the 1s-2p transition energy (cf. the Supplementary Material). When the dielectric constant $\epsilon_s$ is increased, the exciton Bohr radius---and therefore the 1s-2p transition dipole---is enhanced, and at the same time the 1s-2p transition energy is reduced \cite{merkl2019ultrafast}, thus boosting the polarizability. However, the weakening of the Coulomb potential $V^{hh}_{\bm{q}}$ with increasing dielectric constant dominates over the enhancement of the polarizability. This results in a less attractive (negative) Coulomb-hole interaction for high dielectric constants. As a consequence, the Coulomb-hole term counteracts the dominating repulsive dipole-dipole interactions more strongly for free-standing TMD heterostructures.  \\
\begin{figure}[t!]
    \centering
    \includegraphics[width=\columnwidth]{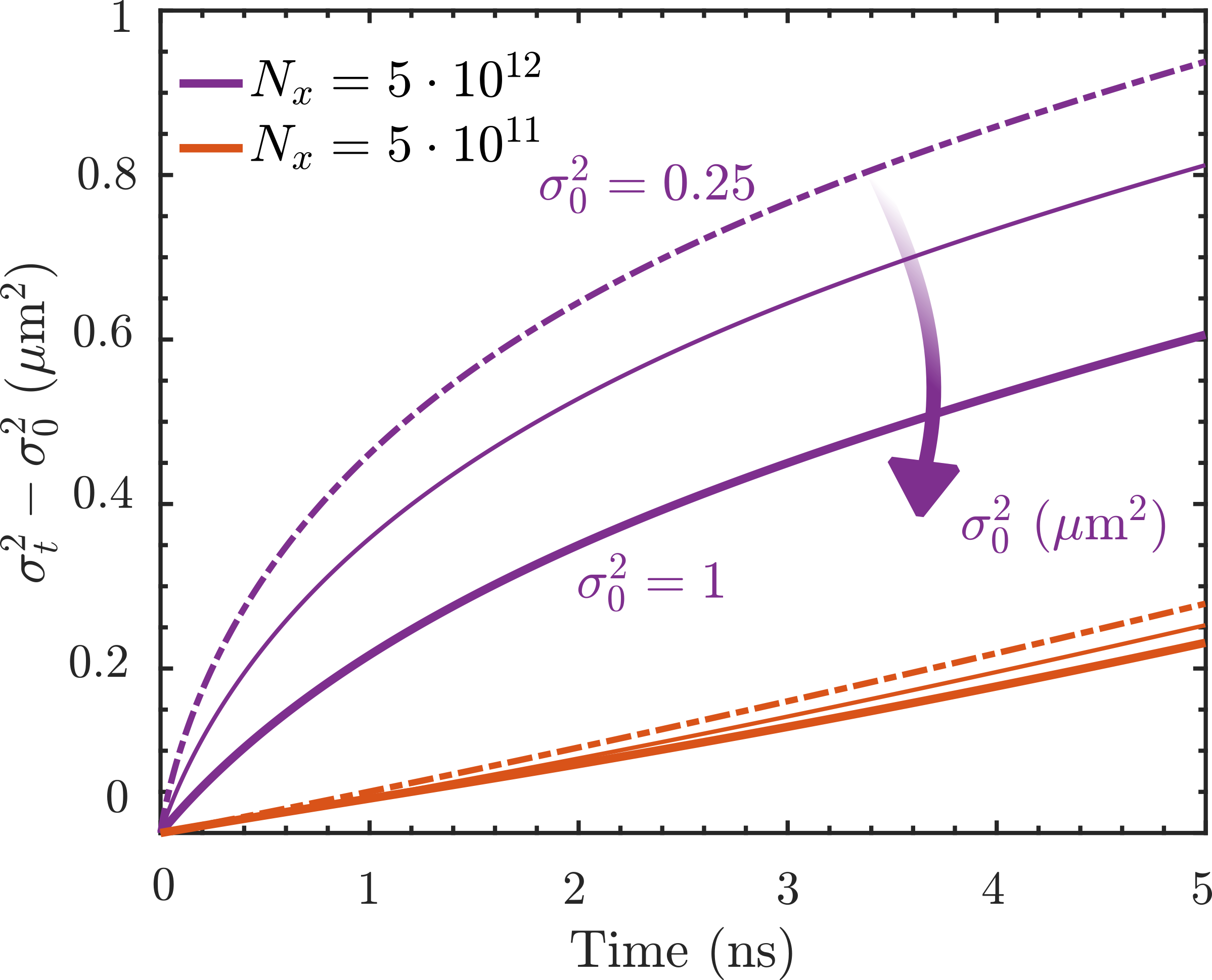}
    \caption{Impact of initial laser spot size on interlayer exciton transport. The exciton density is initialized as $n(x,y,0)=\frac{N_x}{\pi \sigma_0^2} \mathrm{exp}(-(x^2+y^2)/\sigma_0^2)$ such that the total number of excitons $N_x$ is kept constant when varying the spot size, $\sigma_0^2$. For two different  $N_x$, we study different spot sizes  $\sigma_0^2=0.25$, 0.5 and 1  (in units of $\mu\mathrm{m}^2$). Increasing the spot size effectively results in a smaller initial exciton density, making exciton-exciton interaction less important and anomalous exciton diffusion less enhanced.}
    \label{spotsize}
\end{figure}
\newline 
\newline 
\textbf{Excitation spot size:}
Here, we study the impact of the laser spot size on the interlayer exciton transport. By initalizing the exciton density as $n(x,y,0)=\frac{N_x}{\pi \sigma_0^2}\mathrm{exp}(-(x^2+y^2)/\sigma_0^2)$ and keeping the total number of excitons, $N_x$, constant (i.e. considering a constant laser pump power), we investigate  the change in the variance for different  initial laser spots with $\sigma_0^2=0.25, 0.5, 1 \ \mu\mathrm{m}^2$, cf. Fig. \ref{spotsize}. For a relatively low number of excitons, $N_x=5\cdot 10^{11}$ (orange lines), we find that the Coulomb-induced drift is less important and conventional diffusion dominates the exciton transport. As expected from Fick's law of conventional diffusion, the choice of $\sigma_0^2$ is not expected to significantly affect the propagation. However, as we increase the number of excitons $N_x$ to $N_x=5\cdot 10^{12}$ (purple lines), we observe a much more pronounced effect of the initial spot size. We find a decrease in the effective diffusion coefficient from $D_{\mathrm{eff}}(t_0)\approx{1}$ $\mathrm{cm}^2/\mathrm{s}$ to $D_{\mathrm{eff}}(t_0)\approx{0.5}$ $\mathrm{cm}^2/\mathrm{s}$ at $t_0=0.5$ ns when increasing the spot size from $\sigma_0^2=0.25$ $\mu\mathrm{m}^2$ to $\sigma_0^2=1$ $\mu\mathrm{m}^2$. This can be explained by the fact that increasing the spot size effectively reduces the initial exciton density at the center of the excitation spot. 

\section*{Conclusions}
We have investigated the impact of exciton-exciton interaction on the propagation of excitons in the exemplary MoSe$_2$-WSe$_2$ heterostructure. We find that there is a competition between the repulsive dipole-dipole coupling with exchange interaction and exciton screening. The interplay of these processes gives rise to a net Coulomb-induced drift that accelerates the propagation of interlayer excitons at elevated densities.
Furthermore, we demonstrate how this Coulomb-induced drift and the resulting anomalous exciton propagation can be tuned by changing the interlayer separation (e.g. varying the number of hBN spacers), dielectric environment or laser spot size. The developed approach could be generalized to include also the hybridization between intra- and interlayer excitons allowing for the  investigation of exciton-exciton interactions in any van der Waals heterostructure. Overall, our work contributes to a better microscopic understanding of interlayer exciton transport in technologically promising atomically thin semiconductors.
\section*{Acknowledgments}
We thank Joakim Hagel (Chalmers) for stimulating discussions. This project has received funding from  Deutsche Forschungsgemeinschaft via CRC 1083 (Project No. B09) and the European Unions Horizon 2020 research and innovation programme under grant agreement no. 881603 (Graphene Flagship). 

\end{document}


\renewcommand{\theequation}{S\arabic{equation}}
\renewcommand{\thesection}{\Roman{section}}
\preprint{APS/123-QED}

\title{Supplementary Material for \--- Microscopic origin of anomalous interlayer exciton transport in van der Waals heterostructures}
\title{\textbf{Supplementary Material for} \\ Microscopic origin of anomalous interlayer exciton transport in van der Waals heterostructures}
\author{Daniel Erkensten$^1$, Samuel Brem$^2$, Ra\"ul Perea-Caus\'in$^1$, and Ermin Malic$^{2,1}$}
\affiliation{$^1$Department of Physics, Chalmers University of Technology, 41296 Gothenburg, Sweden}
\affiliation{$^2$Department of Physics, Philipps-Universit{\"a}t Marburg, 35037 Marburg, Germany}

\maketitle 
\date{\today}

\section{Excitonic Hamiltonian} 
To be able to describe exciton-exciton interactions in a two-dimensional semiconductor on a microscopic footing we start from a purely electronic two-band Hamilton operator $H=H_0+H_{c-c}$ consisting of kinetic energy ($H_0$) and contributions from the Coulomb interaction ($H_{c-c}$), with
\begin{equation}
    H_0=\sum_{\lambda \bm{k}}\epsilon^{\lambda}_{\bm{k}}\lambda^{\dagger}_{\bm{k}}\lambda_{\bm{k}} \ , \ \ \ \ H_{c-c}=\frac{1}{2}\sum_{\lambda \lambda' \bm{k}, \bm{k'}, \bm{q}}V^{\lambda \lambda'\lambda'\lambda}_{\bm{q}}\lambda^{\dagger}_{\bm{k+q}}\lambda'^{\dagger}_{\bm{k'}-\bm{q}}\lambda'_{\bm{k'}}\lambda_{\bm{k}} \ , 
\end{equation}
where  $V^{\lambda \lambda'\lambda' \lambda}_{\bm{q}}$ is the Coulomb matrix element and $\epsilon^{\lambda}_{\bm{k}}$ is the energy dispersion of the band $\lambda=c,v$ ($c$ being the conduction band, $v$ being the valence band) and the momentum $\bm{k}$. The operator $\lambda_{\bm{k}}$ ($\lambda^{\dagger}_{\bm{k}}$) annihilates (creates) an electron in the band $\lambda$ with the momentum $\bm{k}$. Note that we restrict ourselves to electrons and holes around the K-point in this work and we treat the band dispersion within the effective mass approximation. Moreover, we introduce the notation $V^{cccc}_{\bm{q}}\equiv V^{cc}_{\bm{q}}$, $V^{vvvv}_{\bm{q}}\equiv V^{vv}_{\bm{q}}$ and $V^{cvvc}_{\bm{q}}\equiv V^{cv}_{\bm{q}}$ for the sake of brevity. 

Now, we consider the equation of motion for the microscopic polarisation $\langle P^{\dagger}_{\bm{k}_1+\bm{Q}, \bm{k}_1}\rangle\equiv \langle c^{\dagger}_{\bm{k}_1+\bm{Q}}v_{\bm{k}_1}\rangle$ obtained from the corresponding Heisenberg equation of motion:
\begin{equation}
\begin{aligned}
    \partial_t\langle P^{\dagger}_{\bm{k}_1+\bm{Q}, \bm{k}_1}\rangle&= \frac{i}{\hbar}(\epsilon^c_{\bm{k}_1+\bm{Q}}-\epsilon^v_{\bm{k}_1}+\sum_{\bm{q}}V^{vv}_{\bm{q}})\langle P^{\dagger}_{\bm{k}_1+\bm{Q}, \bm{k}_1}\rangle-\frac{i}{\hbar}\sum_{\bm{q}}V^{cv}_{\bm{q}}\langle P^{\dagger}_{\bm{k}_1+\bm{Q}+\bm{q}, \bm{k}_1+\bm{q}}\rangle\\
    &\ \ \ \ -\frac{i}{\hbar}\sum_{\bm{k},\bm{q}}V^{cv}_{\bm{q}}(\langle c^{\dagger}_{\bm{k}_1+\bm{Q}+\bm{q}}v_{\bm{k}_1}v_{\bm{k}}v^{\dagger}_{\bm{k-q}}\rangle+\langle c^{\dagger}_{\bm{k}_1+\bm{Q}}v_{\bm{k}_1+\bm{q}}c^{\dagger}_{\bm{k+q}}c_{\bm{k}}\rangle) \\
    & \ \ \ \ +\frac{i}{\hbar}\sum_{\bm{k}, \bm{q}}(V^{cc}_{\bm{q}}\langle c^{\dagger}_{\bm{k}_1+\bm{Q}-\bm{q}}v_{\bm{k}_1}c^{\dagger}_{\bm{k+q}}c_{\bm{k}}\rangle+V^{vv}_{\bm{q}}\langle c^{\dagger}_{\bm{k}_1+\bm{Q}}v_{\bm{k}_1+\bm{q}}v_{\bm{k}}v^{\dagger}_{\bm{k+q}}\rangle )\ . 
    \end{aligned}
    \label{full}
    \end{equation}
 Next, we apply the Hartree-Fock factorization scheme for the electronic operators \cite{kira2006many} and transform to the excitonic basis yielding the following equation of motion:
\begin{equation}
\begin{aligned}
    \partial_t P_{n, \bm{Q}}&=\frac{i}{\hbar}\bigg( (E^{n}_{\bm{Q}}+\sum_{\bm{q}}V^{vv}_{\bm{q}})P_{n, \bm{Q}}+\sum_{\substack{\bm{k}, \bm{k}_1, \bm{q}, \bm{Q}_1\\ i,j,m}}P_{m, \bm{Q-q}}\bigg( N^{ij}_{\bm{Q}_1+\bm{q}, \bm{Q}_1}W^{imnj}_{\bm{q}}\\
   & \ \  +N^{ij}_{\bm{Q}_1+\bm{q}, \bm{Q}_1}\varphi^*_{i, \bm{k}_1+\beta (\bm{Q}_1+\bm{q}-\bm{Q})}\varphi_{j, \bm{k}_1+\beta (\bm{Q}_1+\bm{q}-\bm{Q})+\alpha\bm{q}}\varphi^*_{m, \bm{k}+\bm{k}_1+\alpha\bm{q}}(V^{cv}_{\bm{k}}\varphi_{n, \bm{k}_1}-V^{cc}_{\bm{k}}\varphi_{n, \bm{k}_1+\bm{k}})\\
   & \ \ \ +N^{ij}_{\bm{Q}_1+q, \bm{Q}_1}\varphi^*_{i, \bm{k}_1-\alpha(\bm{Q}_1+\bm{q}-\bm{Q})}\varphi_{j,\bm{k}_1-\alpha(\bm{Q}_1+\bm{q}-\bm{Q})-\beta \bm{q}}\varphi^*_{m,\bm{k}+\bm{k}_1-\beta\bm{q}}(V^{cv}_{\bm{k}}\varphi_{n, \bm{k}_1}-V^{vv}_{\bm{k}}\varphi_{n, \bm{k}_1+\bm{k}})\bigg) \ ,
\end{aligned}
\label{fulleq}
\end{equation}
with $P_{n, \bm{Q}}=\langle X^{\dagger}_{n, \bm{Q}}\rangle$, $N^{ij}_{\bm{Q}, \bm{Q'}}=\langle X^{\dagger}_{i, \bm{Q}}X_{j, \bm{Q'}}\rangle$ and the exciton dispersion $E^n_{\bm{Q}}=E^n+\frac{\hbar^2\bm{Q}^2}{2M}$, $M=m_e+m_h$, $m_e$ ($m_h$) being the electron (valence) band masses and $E^n$ is the excitonic binding energy which is extracted from the Wannier equation \cite{haug2009quantum} along with excitonic wave functions $\varphi_{n, \bm{q}}$, $n=1s,2s...$. Note that all explicitly density-dependent quantities appear in the term proportional to $P_{m, \bm{Q}-\bm{q}}$. We also identified the \emph{direct} exciton-exciton interaction \cite{erkensten} 
\begin{equation}
   W^{imnj}_{\bm{q}}=V^{cc}_{\bm{q}} F_{mn}(\beta\bm{q})F^*_{ji}(\beta \bm{q})+V^{vv}_{\bm{q}}F^*_{nm}(\alpha\bm{q})F_{ij}(\alpha\bm{q})-V^{cv}_{\bm{q}}(F_{mn}(\beta\bm{q})F_{ij}(\alpha\bm{q})+F^*_{nm}(\alpha\bm{q})F_{ji}^*(\beta\bm{q}))
\end{equation}
with the form-factors $F_{ij}(x\bm{q})=\sum_{\bm{q'}}\varphi^*_{i, \bm{q'}+x\bm{q}}\varphi_{j, \bm{q'}}$. The remaining terms proportional to $P_{m, \bm{Q-q}}$ constitute fermionic exchange terms, reflecting exchange of the fermionic constituents of the interacting excitons. To arrive at Eq. \eqref{fulleq}, we expand the pair operator $P^{\dagger}$ in an excitonic basis according to 
\begin{equation}
    P^{\dagger}_{\bm{k}, \bm{k'}}=\sum_{n} X^{\dagger}_{n, \bm{k}-\bm{k'}}\varphi^*_{n, \beta \bm{k}+\alpha\bm{k'}} \ , 
\end{equation}
where $X^{\dagger}_{n, \bm{k}-\bm{k'}}$ are excitonic operators. The coefficients $\alpha=\frac{m_e}{m_h+m_e}$ and $\beta=\frac{m_h}{m_e+m_h}$ with electronic masses are obtained from ab-initio calculations \cite{kormanyos2015k}. Moreover, we made use of the following expansions in order to transfer electronic expectation values to excitonic expectation values \cite{katsch2018theory}: 
\begin{equation}
\langle c^{\dagger}_{\bm{k}}c_{\bm{k'}}\rangle \approx{\sum_{\bm{k''}} \langle P^{\dagger}_{\bm{k}, \bm{k''}}P_{\bm{k'}, \bm{k''}}\rangle} \ ,\langle v_{\bm{k}}v^{\dagger}_{\bm{k'}}\rangle \approx{ \sum_{\bm{k''}} \langle P^{\dagger}_{ \bm{k''}, \bm{k}}P_{\bm{k''}, \bm{k'}}\rangle } \ ,        
\end{equation}
omitting terms which are quadratic and of higher order in pair density. 

Now, we  define a mean-field exciton-exciton interaction according to 
\begin{equation}
    W_{\mathrm{mf}, \bm{q}, \bm{Q}}^{mn}=\sum_{\substack{\bm{Q}_1 \\ i,j}}N^{ij}_{\bm{Q}_1+\bm{q}, \bm{Q}_1}\mathcal{W}^{imnj}_{\bm{q}, \bm{Q}_1, \bm{Q}} \ , 
    \label{s77}
\end{equation}
with 
\begin{gather}
\begin{aligned}
    \mathcal{W}^{imnj}_{\bm{q}, \bm{Q}_1, \bm{Q}}&=W^{imnj}_{\bm{q}}  +\sum_{\bm{k}_1, \bm{k}_2}\varphi^*_{i, \bm{k}_1+\beta (\bm{Q}_1+\bm{q}-\bm{Q})}\varphi_{j, \bm{k}_1+\beta (\bm{Q}_1+\bm{q}-\bm{Q})+\alpha\bm{q}}\varphi^*_{m, \bm{k}+\bm{k}_1+\alpha\bm{q}}(V^{cv}_{\bm{k}}\varphi_{n, \bm{k}_1}-V^{cc}_{\bm{k}}\varphi_{n, \bm{k}_1+\bm{k}})\\
   & +\sum_{\bm{k}_1, \bm{k}_2}\varphi^*_{i, \bm{k}_1-\alpha(\bm{Q}_1+\bm{q}-\bm{Q})}\varphi_{j,\bm{k}_1-\alpha(\bm{Q}_1+\bm{q}-\bm{Q})-\beta \bm{q}}\varphi^*_{m,\bm{k}+\bm{k}_1-\beta\bm{q}}(V^{cv}_{\bm{k}}\varphi_{n, \bm{k}_1}-V^{vv}_{\bm{k}}\varphi_{n, \bm{k}_1+\bm{k}}) \ , 
\end{aligned}
\label{effmatrix}
\end{gather}
such that a commutation between the excitonic polarisation and the Hamiltonian $H_x=H_{x,0}+H^{\mathrm{mf}}_{x-x}$ with
\begin{equation}
  H_{x,0}=\sum_{n, \bm{Q}}\bigg(E^n_{\bm{Q}}+\sum_{\bm{q}}V^{vv}_{\bm{q}}\bigg)X^{\dagger}_{n, \bm{Q}}X_{n, \bm{Q}} \ ,  H^{\mathrm{mf}}_{x-x}=\sum_{m,l, \bm{q}, \bm{Q}_1}W^{ml}_{\mathrm{mf}, \bm{q}, \bm{Q}_1}X^{\dagger}_{m, \bm{Q}_1-\bm{q}}X_{l, \bm{Q}_1} \ , 
  \label{hamo}
\end{equation}
reproduces Eq. \eqref{fulleq} when treating $X^{(\dagger)}$ as bosonic operators. In this work, we make use of the simple case $\bm{q}=\bm{Q}=0$ when evaluating the mean-field exciton-exciton interaction. Moreover, we only consider interactions between $m=n=1s$ excitons (and can consequently drop the excitonic indices) yielding the simplified exciton-exciton interaction
\begin{equation}
    \lim_{\bm{q}, \bm{Q}\rightarrow 0}W_{\mathrm{mf}, \bm{q}, \bm{Q}}\approx{n_x (g_{d-d}+g_{x-x})} \ , 
\end{equation}
where $n_x\equiv \frac{1}{A}\sum_{\bm{Q}_1} N_{\bm{Q}_1, \bm{Q}_1}$ is the exciton density, $A$ being the crystal area, and 
\begin{equation}
    g_{d-d}\equiv A(V^{cc}_{\bm{q}}+V^{vv}_{\bm{q}}-2V^{cv}_{\bm{q}})\delta_{\bm{q}, 0} \ ,  
\end{equation}
and 
\begin{equation}
    g_{x-x}\equiv A\sum_{\bm{k}, \bm{k'}}\bigg(\varphi^*_{ \bm{k'}}\varphi_{\bm{k'}}\varphi^*_{\bm{k}+\bm{k'}}(V^{cv}_{\bm{k}}\varphi_{n, \bm{k'}}-V^{vv}_{\bm{k}}\varphi_{\bm{k'}+\bm{k}})+\varphi^*_{ \bm{k'}}\varphi_{\bm{k'}}\varphi^*_{\bm{k}+\bm{k'}}(V^{cv}_{\bm{k}}\varphi_{n, \bm{k'}}-V^{cc}_{\bm{k}}\varphi_{\bm{k'}+\bm{k}})\bigg) \ . 
    \label{gxx}
\end{equation}
Here, we note that $g_{d-d}$ stems from \emph{direct} exciton-exciton interaction and $g_{x-x}$ is  due to the \emph{exchange} of fermionic constituents \cite{exchange}. Note that we have excluded the electron-hole exchange interaction in the derivation above. As this interaction scales with the optical matrix element or transition dipole element\cite{kira2006many}, it can be safely neglected \cite{eduardo} when considering interactions between interlayer excitons due to the large separation between electrons and holes. In the intralayer case we perform an analogous calculation as above and find a repulsive correction term to the exchange term reading \cite{katsch2018theory}
\begin{equation}
    g_{x-x}|_{\mathrm{e-h\ exch.}}=2A\sum_{\bm{k}, \bm{k'}}\bar{V}^{cv}_{\bm{k}}|\varphi_{\bm{k}}|^2|\varphi_{\bm{k'}}|^2 \ , 
\end{equation}
with $\bar{V}^{cv}_{\bm{q}}=V_{\bm{q}} |\bm{q}\cdot \bm{M}^{cv}|^2$, where the optical matrix element $|M^{cv}|^2$ is obtained as $|M^{cv}|^2=2a_0^2 t^2/E_g^2$ projected on the normalized Jones vectors, $E_g$ being the single particle bandgap, $a_0$ being the lattice constant and $t$ being an effective hopping integral \cite{PhysRevLett.124.257402, PhysRevLett.108.196802}. The parameters used to compute the optical matrix element in monolayer WSe$_2$ are provided in Table I. 
Generally, considering long-range interactions between intralayer excitons (X) in a monolayer it holds that $V^{cc}_{\bm{q}}\approx{V^{vv}_{\bm{q}}\approx{V^{cv}_{\bm{q}}}}$ and consequently $g^{(X)}_{d-d}\approx{0}$. However, considering interlayer excitons (IX) it holds that all Coulomb matrix elements are different as the matrix elements are screened differently depending on the band configuration. In particular, introducing the layer indices $l, \bar{l}$ and assuming that the electron resides in layer $l$ and the hole resides in layer $\bar{l}$ one finds that 
\begin{equation}
    V^{c_{l} c_{l}}_{\bm{q}}=\frac{W_{\bm{q}}}{\epsilon^{l l}_{\bm{q}}} \ ,  V^{v_{\bar{l}} v_{\bar{l}}}_{\bm{q}}=\frac{W_{\bm{q}}}{\epsilon^{\bar{l} \bar{l}}_{\bm{q}}}  \ ,  V^{c_{l} v_{\bar{l}}}_{\bm{q}}=\frac{W_{\bm{q}}}{\epsilon^{l \bar{l}}_{\bm{q}}} \ , 
\end{equation}
where $W_{\bm{q}}=\frac{e_0^2 }{2\epsilon_0 A|\bm{q}|}$ is the bare Coulomb interaction, and $\epsilon^{l l}_{\bm{q}}, \epsilon^{\bar{l} \bar{l}}_{\bm{q}}, \epsilon^{l \bar{l}}_{\bm{q}}$ are layer-dependent dielectric functions taking the screening from the surrounding TMD layers and substrate into account. These functions can be obtained by solving the two-dimensional Poisson equation for two homogenous slabs and read \cite{samuelthesis}
\begin{align}
\epsilon^{l\bar{l}}_{\mathbf{q}}=\kappa g^{l}_{\mathbf{q}}g^{\bar{l}}_{\mathbf{q}}f_{\mathbf{q}} \ , \epsilon^{l l}_{\mathbf{q}}=\frac{\kappa g^{\bar{l}}_{\mathbf{q}}f_{\mathbf{q}}}{\cosh(\delta_{\bar{l}}{|\mathbf{q}|}/2)h^{l}_{\mathbf{q}}} \ , \epsilon^{\bar{l} \bar{l}}_{\mathbf{q}}=\frac{\kappa g^{l}_{\mathbf{q}}f_{\mathbf{q}}}{\cosh(\delta_{l}{|\mathbf{q}|}/2)h^{\bar{l}}_{\mathbf{q}}}
\label{dilfunc}
\end{align}
with the abbreviations 
\begin{align}
f_{\mathbf{q}}&=1+\frac{1}{2}\left((\frac{\kappa_l}{\kappa}+\frac{\kappa}{\kappa_l})\tanh(\delta_l|\mathbf{q}|)+(\frac{\kappa_{\bar{l}}}{\kappa}+\frac{\kappa}{\kappa_{\bar{l}}})\tanh(\delta_{\bar{l}}|\mathbf{q}|)+(\frac{\kappa_l}{\kappa_{\bar{l}}}+\frac{\kappa_{\bar{l}}}{\kappa_l})\tanh(\delta_l |\mathbf{q}|)\tanh(\delta_{\bar{l}} |\mathbf{q}|)\right) \\
h_{\mathbf{q}}^l&=1+\frac{\kappa}{\kappa_l}\tanh(\delta_l|\mathbf{q}|)+\frac{\kappa}{\kappa_{\bar{l}}}\tanh(\delta_{\bar{l}}|\mathbf{q}|/2)+\frac{\kappa_l}{\kappa_{\bar{l}}}\tanh(\delta_l|\mathbf{q}|)\tanh(\delta_{\bar{l}}|\mathbf{q}|/2) \\
g_{\mathbf{q}}^l&=\frac{\cosh(\delta_l|\mathbf{q}|)}{\cosh(\delta_{\bar{l}}|\mathbf{q}|/2)}\left(1+\frac{\kappa}{\kappa_l}\tanh(\delta_l|\mathbf{q}|/2)\right)^{-1} .
\end{align}
Note that $f_{\bm{q}}$ is symmetric with respect to $l$ and $\bar{l}$ and that the expressions for $h^{\bar{l}}_{\bm{q}}$ and $g^{\bar{l}}_{\bm{q}}$ are obtained from interchanging $l\leftrightarrow \bar{l}$. The parameters are $\kappa_l=\sqrt{\varepsilon_{\parallel}^l\varepsilon_{\perp}^l}$ and $\kappa$ for the dielectric background (computed as the algebraic average between the dielectric constant of the substrate below and above the TMD layers respectively), $\alpha_l=\sqrt{\varepsilon^l_{\parallel}/\varepsilon^l_{\perp}}$, $\delta_l=\alpha_ld_l$ with the layer thickness $d_l$. The in-plane ($\varepsilon_{||}$), out-of-plane ($\varepsilon_{\perp}$) components of the dielectric tensors of the TMDs and layer thicknesses are taken from ab-initio calculations and are summarized in Table I in Section IV \cite{laturia2018dielectric}. Now, by expanding the dielectric functions for \emph{small} momentum $\bm{q}$ one deduces that the direct interlayer exciton-exciton interaction is given  by 
\begin{equation}
    g_{d-d}^{(IX)}=\frac{e_0^2}{2\epsilon_0 }(\frac{d_l }{\varepsilon_{\perp}^l}+\frac{d_{\bar{l}} }{\varepsilon_{\perp}^{\bar{l}}})+\mathcal{O}(\bm{q}) \ , 
    \label{gdd}
\end{equation}
which is precisely a classical dipole-dipole interaction (justifying the subscript "d-d"). The total interlayer exciton-exciton interaction is given by $g_{IX}=g_{x-x}+g_{d-d}$, with $g_{d-d}$ dominating over $g_{x-x}$ as is shown in the main text. 

Note that in the main manuscript we include hBN spacers between the TMD layers forming the heterostructure. This requires us to modify the dielectric screening functions above, with the generalized expressions being provided in \cite{ovesen2019interlayer}. A similar analysis as above can be carried out in this case yielding the dipole-dipole interaction as a function of the number of hBN spacers, $n$:
\begin{equation}
    g_{d-d}^{(IX)}(n)=\frac{e_0^2}{2\epsilon_0 }(\frac{d_l }{\varepsilon_{\perp}^l}+\frac{d_{\bar{l}} }{\varepsilon_{\perp}^{\bar{l}}}+\frac{n d_{\mathrm{hBN}}}{\kappa_{\mathrm{hBN}}}) \ , 
\end{equation}
where $d_{\mathrm{hBN}}=0.3$ nm is the thickness of the hBN layer and $\kappa_{\mathrm{hBN}}\approx 4.5$ \cite{hbnref}. The corresponding interlayer exchange contribution $g_{x-x}^{(IX)}$ is more difficult to analyze due to the convolution of excitonic wave functions entering $g_{x-x}^{(IX)}$. However, we note that the electron-hole Coulomb matrix element $V^{cv}$ rapidly decreases with the TMD layer separation and consequently $g_{x-x}$ becomes more attractive when increasing the number of hBN spacers. This has been observed previously in literature, in particular in the context of coupled quantum wells \cite{kyrienko}. Having discussed dielectric screening of the intra- and interlayer Coulomb matrix elements, we also note that excitons are screened by \emph{themselves} at elevated densities \cite{ropke1979influence, ropke1981green}. This requires us to take $\epsilon_{\bm{q}}\rightarrow \tilde{\epsilon}_{\bm{q}}=\epsilon_{\bm{q}}\epsilon_{exc, \bm{q}}(n_x)$ in our Coulomb matrix elements, where $\epsilon_{exc, \bm{q}}(n_x)$ describes the density-dependent screening function due to the presence of other excitons in the heterostructure. The particular form of $\epsilon_{exc, \bm{q}}(n_x)$ is discussed in the next section. Moreover, the band gap renormalizations due to a filled valence band, $\sum_{\bm{q}} V^{vv}_{\bm{q}}$, entering Eq. \eqref{fulleq} is modified to $\Sigma_{CH} +\sum_{\bm{q}} V^{vv}_{\bm{q}}$, introducing the density-dependent \emph{Coulomb-hole} contribution $\Sigma_{CH}=\bigg(\frac{V^{vv}_{\bm{q}}}{\epsilon_{exc, \bm{q}}}-V^{vv}_{\bm{q}}\bigg)$ \cite{hedin1965new, hybertsen1985first}. The remaining term ($\sum_{\bm{q}} V^{vv}_{\bm{q}}$) is absorbed in the single-particle band gap in the ground state dispersion. Note that self-consistent calculations of the Coulomb-hole based on exciton Greens functions predict that valence and conduction bands are equally screened such that $\Sigma_{CH}=\frac{1}{2}\sum_{\lambda=c,v} (V^{\lambda\lambda}_{s, \bm{q}}-V^{\lambda\lambda}_{\bm{q}})$ \cite{PhysRevB.98.035434}, coinciding with our result when $V^{vv}_{\bm{q}}\approx{V^{cc}_{\bm{q}}}$, which holds assuming that the electrons and holes are well-localized around the high-symmetry points.

Finally, we provide the equation of motion for exciton polarisation only including $n=1s$ states with $\bm{Q}=0$. Then, Eq. \eqref{fulleq} reduces to the following within the \emph{COHSEX} approximation \cite{hedin1965new, hybertsen1985first}
\begin{equation}
    \partial_t P_{1s}=\frac{i}{\hbar}\bigg(E^{1s}+\Sigma_{CH}+(g_{d-d}+\bar{g}_{x-x})n_x\bigg)P_{1s} \ , 
\end{equation}
where we introduced the \emph{screened} exchange term $\bar{g}_{x-x}$ ($g_{x-x}$ with $\epsilon_{\bm{q}}\rightarrow \epsilon_{\bm{q}}\epsilon_{exc, \bm{q}}$). The dipole-dipole contribution is unscreened to lowest order in momentum (as $\epsilon_{exc,\bm{q}=0}=1$).

Although the derivation of the mean-field Hamiltonian (Eq. \eqref{hamo}) presented here is novel and done within the exciton density matrix formalism, exciton-exciton interactions have been also studied in the past via exciton Greens functions \cite{boldt1985many, may1985many}. In particular, we note that the Hartree-Fock self-energy introduced in Ref. \cite{boldt1985many, may1985many} display a perfect analogy to the energy renormalizations $(g_{d-d}+g_{x-x})n_x$ derived in this work. However, we  disregard the impact of higher-order correlations. 

\section{Exciton screening}
At elevated densities below the Mott transition, most electron-hole pairs are bound as excitons and similar to how electrons and holes are screened by additional charges, also excitons are screened by other excitons. Here, we will derive the excitonic screening due to interlayer excitons in a heterobilayer. To begin with, we consider the heterostructure as two homogenous slabs separated by a distance $d$ in vacuum. We put a test charge in $z=0$ and allow for an \emph{induced charge} $\rho^l(\bm{r})$ in layer $l$. The induced charge depends on both the potential in layer $l$, $W_{\bm{q}, z=0}$ and the potential in the other layer $\bar{l}$, $W_{\bm{q}, z=d}$ and can be expressed in momentum space as
\begin{equation}
    \rho^l_{\bm{q}}=W_{\bm{q}, 0}A^{l\bar{l}}_{\bm{q}}-W_{\bm{q}, d}B^{l\bar{l}}_{\bm{q}} \ , 
\end{equation}
treated in RPA \cite{kira2006many} where 
\begin{equation}
    A^{l\bar{l}}_{\bm{q}}=\sum_{\mu\nu}\bigg(|(F_{\mu\nu}(\beta^{l\bar{l}}\bm{q})|^2\Pi^{l\bar{l}}_{\mu\nu}(\bm{q})+|F_{\mu\nu}(-\alpha^{\bar{l}l}\bm{q})|^2\Pi^{\bar{l}l}_{\mu\nu}(\bm{q})\bigg) \ , 
    \label{Aeq}
\end{equation}
and 
\begin{equation}
     B^{l\bar{l}}_{\bm{q}}=\sum_{\mu\nu}\bigg(F_{\mu\nu}(\beta^{l\bar{l}}\bm{q})F_{\mu\nu}(-\alpha^{\bar{l}l}\bm{q})^*\Pi^{l\bar{l}}_{\mu\nu}(\bm{q})+F_{\mu\nu}(\beta^{\bar{l}l}\bm{q})^*F_{\mu\nu}(-\alpha^{\bar{l}l}\bm{q})\Pi^{\bar{l}l}_{\mu\nu}(\bm{q})\bigg) \ . 
     \label{Beq}
\end{equation}
Here, we introduced the \emph{excitonic polarizability} in the static limit
\begin{equation}
    \Pi^{l\bar{l}}_{\mu\nu}(\bm{q})=\sum_{\bm{Q}}\frac{N^{\nu}_{l\bar{l}, \bm{Q}}-N^{\mu}_{l\bar{l}, \bm{Q+q}}}{E^{\nu}_{l\bar{l},\bm{Q}}-E^{\mu}_{l\bar{l}, \bm{Q+q}}} \ , 
\end{equation}
where $\mu,\nu=1s,2p,2s...$ is the exciton state and $N^{\mu}_{l\bar{l}, \bm{Q}}$ is the interlayer exciton occupation. At cryogenic temperatures and elevated densities, at which the exciton distribution is strongly peaked around $\bm{Q}=0$, the polarizability is directly proportional to the exciton density, $n_x$. Note that we assume that the occupation of intralayer excitons is negligible, which holds due to the large type-II band alignment in the considered MoSe$_2$-WSe$_2$ heterostructure \cite{gillen2018interlayer}. Now, we solve for the intralayer and interlayer potentials, $W_{\bm{q}, 0}$ and $W_{\bm{q}, d}$ the following Poisson equation:
\begin{equation}
    \partial_z^2 W_{\bm{q}, z}-q^2W_{\bm{q}, z}=e\delta(z)+\rho^{(1)}_{\bm{q}}\delta(z)+\rho^{(2)}_{\bm{q}}\delta(z-d) \ ,
    \label{pos}
\end{equation}
where we have applied a Fourier-transformation in the in-plane direction such that $(x,y)\rightarrow (q_x, q_y)$ and defined $l\equiv 1$, $\bar{l}=2$. 
The interlayer ($W_{q,z=d}\equiv \frac{e^2}{2\epsilon_0 |\bm{q}|\epsilon_{inter, exc}(\bm{q})}$) and intralayer ($W_{q,z=0} \equiv \frac{e^2}{2\epsilon_0 |\bm{q}|\epsilon_{intra, exc}(\bm{q})}$) potentials can be found by solving Eq. \eqref{pos}, resulting in the dielectric functions 
\begin{equation}
\epsilon_{intra, exc}(\bm{q})=\frac{\bigg(B^{12}_{\bm{q}} B^{21}_{\bm{q}} -A^{12}_{\bm{q}}A^{21}_{\bm{q}}+\mathrm{e}^{2dq}[-B^{12}_{\bm{q}} B^{21}_{\bm{q}}+(A^{12}_{\bm{q}}-2q)(A^{21}_{\bm{q}}-2q)]+2(B^{12}_{\bm{q}}+B^{21}_{\bm{q}})\mathrm{e}^{dq}q\bigg) }{2q(A^{21}_{\bm{q}}+\mathrm{e}^{2dq}(2q-A^{21}_{\bm{q}}))} \ , 
\label{excintra}
\end{equation}
and 
\begin{equation}
\epsilon_{inter, exc}(\bm{q})=\frac{\bigg(B^{12}_{\bm{q}} B^{21}_{\bm{q}} -A^{12}_{\bm{q}}A^{21}_{\bm{q}}+\mathrm{e}^{2dq}[-B^{12}_{\bm{q}} B^{21}_{\bm{q}}+(A^{12}_{\bm{q}}-2q)(A^{21}_{\bm{q}}-2q)]+2(B^{12}_{\bm{q}}+B^{21}_{\bm{q}})\mathrm{e}^{dq}q\bigg) }{4q(q-B^{21}_{\bm{q}}\sinh(dq))} \ , 
\end{equation}
where the functions $A^{l\bar{l}}_{\bm{q}}$ and $B^{l\bar{l}}_{\bm{q}}$ are provided by Eq. \eqref{Aeq} and Eq. \eqref{Beq} respectively. By considering the limit $qd<<1$ (corresponding effectively to the case of a monolayer) we retrieve a particularly simple expression for the dielectric functions commonly employed in literature for excitonic screening in monolayers \cite{steinhoff2017exciton} or three-dimensional electron-hole plasmas \cite{ropke1979influence}:
\begin{equation}
\epsilon_{inter, exc}(\bm{q})=\epsilon_{intra, exc}(\bm{q})=1-W_{\bm{q}}\sum_{\mu\nu}\tilde{\Pi}^{l\bar{l}}_{\mu\nu}(\bm{q}) \ , 
\end{equation}
with $\tilde{\Pi}^{l\bar{l}}_{\mu\nu}(\bm{q})=\Pi^{l\bar{l}}_{\mu\nu}(\bm{q})|F_{\mu\nu}(\beta^{l\bar{l}} \bm{q})-F_{\mu\nu}(-\alpha^{l\bar{l}} \bm{q})|^2$. Here, the bare Coulomb potential $W_{\bm{q}}$ enters. To include the screening from the dielectric environment and the heterostructure we take $W_{\bm{q}}\rightarrow V^{ll'}_{\bm{q}}=\frac{W_{\bm{q}}}{\epsilon^{ll'}_{\bm{q}}}$ with the dielectric functions given by \eqref{dilfunc}. Note also from the expression above that the excitonic screening depends on the difference in excitonic form factors, reflecting the neutral character and weak polarizability of 1s excitons. These form factors can be expanded for small $q$ and related to the in-plane exciton dipole moment according to $|F_{\mu\nu}(\beta^{l\bar{l}}\bm{q})-F_{\mu\nu}(-\alpha^{l\bar{l}}\bm{q})|^2\approx{(\bm{q}\cdot\bm{d}_{\mu\nu})^2} $, where $\bm{d}_{\mu\nu}=\int d^2\bm{r} \varphi^*_{\mu}(\bm{r})\bm{r}\varphi_{\nu}(\bm{r})$ and notably the in-plane exciton dipole moment of 1s excitons is vanishing as the ground state real space exciton wave functions $\varphi_{1s}(\bm{r})$ are even \cite{merkl2019ultrafast}. However, we include higher-order states $\nu=2p,3p,4p$ in the calculation, such that the transition dipole moment of excitons is non-vanishing and the exciton screening is enhanced. For the dominating 1s-2p transition and in the small $\bm{q}$ limit we can analytically evaluate the exciton polarizability according to 
\begin{equation}
    \tilde{\Pi}_{1s-2p}(\bm{q})\approx{\frac{n_x (\bm{q}\cdot \bm{d}_{1s-2p})^2}{\Delta E_{1s,2p}}} \ , 
\end{equation}
where $\Delta E_{1s,2p}$ is the 1s-2p transition energy and $n_x$ is the density of 1s excitons (assuming the occupation of $p$-states to be negligible). This expression holds for both intra- and interlayer excitons. 

\begin{figure}
    \centering
    \includegraphics[width=0.75\columnwidth]{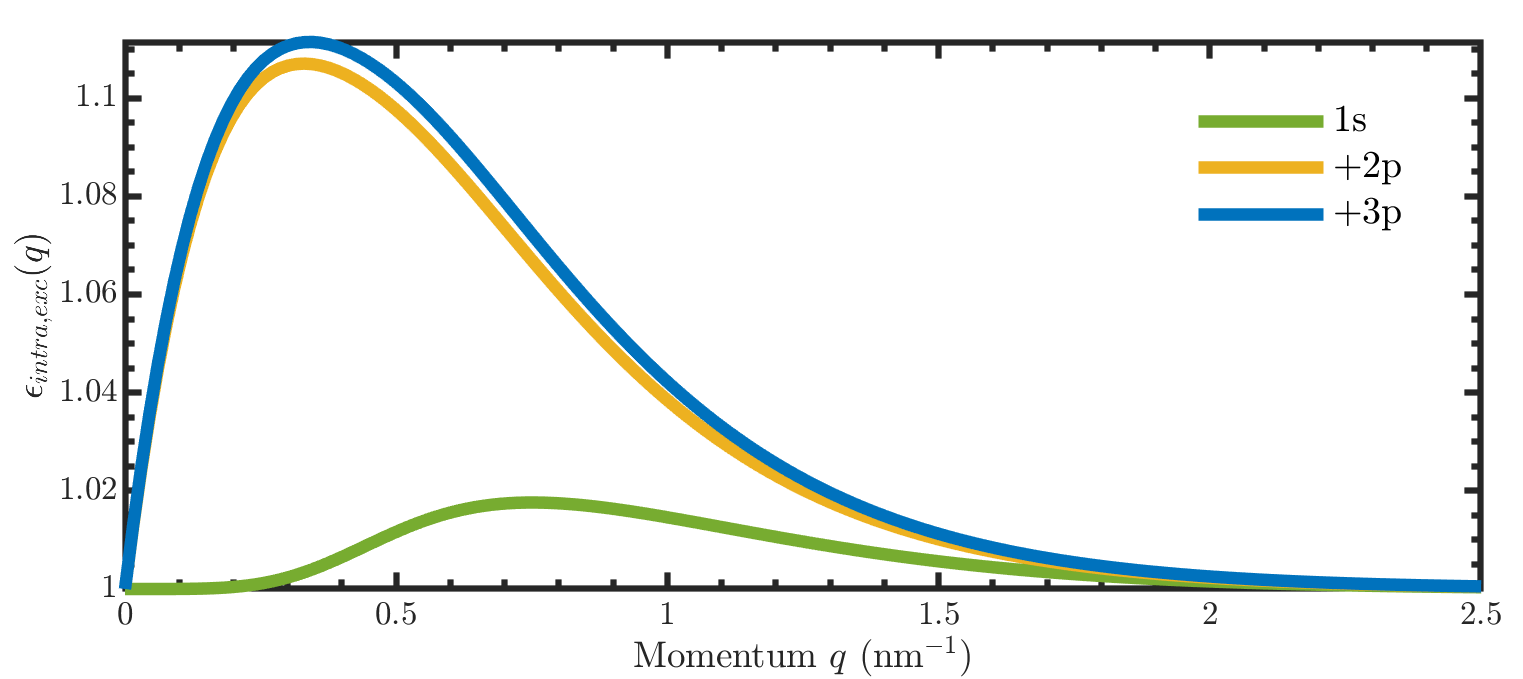}
    \caption{Exciton screening $\epsilon_{intra, exc}(q)$ in MoSe$_2$-hBN-WSe$_2$ separating contributions stemming from $s$ and $p$-states at the exciton density $n_x=10^{12}$ $\mathrm{cm}^{-2}$. As the intra- and interlayer screening approximately scale with the exciton transition dipole moment, the screening is dominated by 1s-2p exciton transitions.}
    \label{screenp}
\end{figure}

In Fig. \ref{screenp} we illustrate the full momentum-dependent intralayer exciton screening (Eq. \eqref{excintra}) for the exemplary MoSe$_2$-hBN-WSe$_2$ heterotrilayer  and separate contributions stemming from 1s-1s,  1s-2p and 1s-3p transitions. We fix the exciton density to  $n_x=10^{12} \mathrm{cm}^{-2}$, temperature to $T=4.6$ K and set $d=d_{\mathrm{hBN}}+d_{\mathrm{TMD}}$ with the thickness of the hBN spacer, $d_{\mathrm{hBN}}=0.3$ nm, and TMD layers $d_{\mathrm{TMD}}=0.65$ nm. The excitonic wave functions entering the form factors are obtained from the Wannier equation \cite{ovesen2019interlayer, kira2006many}. We  neglect $1s-np$ transitions for $n>3$ due to the small excitonic wave function overlap of 1$s$ and higher-lying $p$-states.  

\section{Drift-diffusion equation}
In this section, we make use of the Hamiltonian derived in Section I to find the corresponding drift-diffusion equation for interlayer excitons. The spatiotemporal dynamics of excitons can be accessed via the equation of motion for the exciton Wigner function $N_{\bm{Q}}(\bm{r})=\sum_{\bm{q}} \mathrm{e}^{i\bm{q}\cdot\bm{r}} N_{\bm{Q}+\bm{q}/2, \bm{Q}-\bm{q}/2}$, where $N_{\bm{Q}+\bm{q}/2, \bm{Q}-\bm{q}/2}=\langle X^{\dagger}_{\bm{Q}+\bm{q}/2}X_{\bm{Q}-\bm{q}/2}\rangle$ \cite{hess1996spatio, rosati2020negative}. The strategy is therefore to \textbf{i)} compute the equation of motion for the off-diagonal quantity $N_{\bm{Q}, \bm{Q'}}=\langle X^{\dagger}_{\bm{Q}} X_{\bm{Q'}}\rangle$ and \textbf{ii)} Fourier-transform the result to get an equation of motion in the Wigner representation. To simplify the calculations we drop the dependence on exciton indices considering only interactions between 1s states in the matrix element. Within this approximation we find
\begin{equation}
  \dot{N}_{\bm{Q}, \bm{Q'}}=\frac{i}{\hbar}(E_{\bm{Q}}-E_{\bm{Q'}})N_{\bm{Q}, \bm{Q'}}-\frac{i}{\hbar}\sum_{\bm{q}}(\mathcal{E}_{\bm{Q'}, \bm{q}}N_{\bm{Q}, \bm{q}}-\mathcal{E}_{\bm{q}, \bm{Q}}N_{\bm{q}, \bm{Q'}}) \ , 
\end{equation}
with $\mathcal{E}_{\bm{Q}, \bm{Q'}}=\sum_{\bm{Q}_1}N_{\bm{Q}_1+\bm{Q'}-\bm{Q}, \bm{Q}_1}\mathcal{W}_{\bm{Q'-Q}, \bm{Q'}, \bm{Q}_1}$, where $\mathcal{W}$ is defined in Eq. \eqref{effmatrix} (here omitting exciton indices as only interactions between 1$s$ states are considered). Now, turning to step \textbf{ii)} we take $\bm{Q}\rightarrow \bm{Q}+\bm{q}/2$ and $\bm{Q'}\rightarrow \bm{Q}-\bm{q}/2$, multiply the expression above by $\sum_{\bm{q}}\mathrm{e}^{i\bm{q}\cdot \bm{r}}$ and express the appearing expectation values in term of Wigner functions. The contributions stemming from exciton-exciton interactions read 
\begin{equation}
    \dot{N}_{\bm{Q}}(\bm{r})=-\frac{i}{\hbar}\int d\bm{r'}\sum_{\bm{q}, \bm{q'}}\mathrm{e}^{i\bm{q'}\cdot\bm{r'}}\bigg( \mathcal{E}_{\bm{Q}+\bm{q'}/2+\bm{q}/2, \bm{Q}+\bm{q'}/2-\bm{q}/2}-\mathcal{E}_{\bm{Q}-\bm{q'}/2-\bm{q}/2, \bm{Q}-\bm{q'}/2+\bm{q}/2}\bigg)N_{\bm{Q}-\bm{q}/2}(\bm{r}-\bm{r'}) \ .  
\end{equation}
Here, we note the striking analogy to the purely electronic case, which has been treated in \cite{hess1996maxwell}. In the following, we apply a Taylor expansion with respect to both space ($\bm{r}$) and momentum ($\bm{Q}$)  yielding to first order 
\begin{equation}
    \dot{N}_{\bm{Q}}(\bm{r})\approx{\frac{1}{\hbar}(\frac{\partial \mathcal{E}(\bm{r}, \bm{Q})}{\partial \bm{r}}\frac{\partial N(\bm{Q}, \bm{r})}{\partial \bm{Q}}-\frac{\partial \mathcal{E}(\bm{r}, \bm{Q})}{\partial \bm{Q}}\frac{\partial N(\bm{Q}, \bm{r})}{\partial \bm{r}}}) \ , 
\end{equation}
introducing the spatially dependent energy renormalization $\mathcal{E}(\bm{r}, \bm{Q})=\Phi(\bm{r})+\Delta\mathcal{E}(\bm{r}, \bm{Q})$. Here, we defined 
\begin{equation}
    \Phi(\bm{r})\equiv \sum_{\bm{q}, \bm{Q}_1}\mathrm{e}^{i\bm{q}\cdot\bm{r}}N_{\bm{Q}_1-\bm{q}/2, \bm{Q}_1+\bm{q}/2}W_{\bm{q}}=\int d\bm{r'}n(\bm{r'})W(\bm{r}-\bm{r'}) \ , 
\end{equation}
and 
\begin{align}
    \Delta \mathcal{E}(\bm{r}, \bm{Q})&=\sum_{\bm{q}, \bm{Q}_1, \bm{k}, \bm{k}_1}\mathrm{e}^{i\bm{q}\bm{r}}N_{\bm{Q}_1-\bm{q}/2, \bm{Q}_1+\bm{q}/2}\bigg(\varphi^*_{\bm{k}_1+\beta (\bm{Q}_1-\bm{Q})}\varphi_{\bm{k}_1+\beta (\bm{Q}_1-\bm{Q})-\alpha\bm{q}}\varphi^*_{\bm{k}-\alpha\bm{q}}(V^{cv}_{\bm{k}-\bm{k}_1}\varphi_{\bm{k}_1}-V^{cc}_{\bm{k}-\bm{k}_1}\varphi_{\bm{k}})\\
   & \ \ \ +\varphi^*_{\bm{k}_1-\alpha(\bm{Q}_1-\bm{Q})}\varphi_{\bm{k}_1-\alpha(\bm{Q}_1-\bm{Q})+\beta \bm{q}}\varphi^*_{\bm{k}+\beta\bm{q}}(V^{cv}_{\bm{k}-\bm{k}_1}\varphi_{\bm{k}_1}-V^{vv}_{\bm{k-k_1}}\varphi_{\bm{k}})\bigg)\ .\nonumber  
\end{align}
Now, we employ a small momentum approximation of the exciton-exciton interaction, i.e. consider the relevant momenta for cold exciton distributions and set $\bm{q}\approx{\bm{Q}}\approx{0}$. This procedure has previously been employed in the context of quantum diffusion in coupled quantum wells \cite{ivanov2002quantum}. Then it directly follows that $\mathcal{E}(\bm{r}, \bm{Q})\approx{\mathcal{E}(\bm{r}, 0)}\approx{n(\bm{r})g_{IX}}$, where $n(\bm{r})=\frac{1}{A}\sum_{\bm{Q}_1} N_{\bm{Q}_1, \bm{Q}_1}$ being the spatially dependent exciton density, $g_{IX}=g_{d-d}+g_{x-x}$ with the exchange and dipole-dipole  interactions as defined in Eq.\eqref{gxx} and \eqref{gdd}. Then, a particularly simple equation for the Wigner function can be obtained (now including also the free part)
\begin{equation}
    \dot{N}_{\bm{Q}}(\bm{r})=-v_{\bm{Q}}\nabla_{\bm{r}}N_{\bm{Q}}(\bm{r})- \frac{g_{IX}v_{\bm{Q}}}{k_B T}\cdot\nabla_{\bm{r}}n(\bm{r})N_{\bm{Q}}(\bm{r})(1+N_{\bm{Q}}(\bm{r})) \ , 
\end{equation}
when assuming a Bose distribution in momentum space for excitons and introducing the exciton velocity $v_{\bm{Q}}=\frac{\hbar\bm{Q}}{M}$, $M$ being the total exciton mass. We recognize the equation above as a Boltzmann transport equation, which by invoking the relaxation time approximation and upon summation over $\bm{Q}$ is transformed to a continuity equation $\partial_t n(\bm{r},t)+\nabla\cdot \bm{J}_{\mathrm{exc}}=0$, with the excitonic flux density $\bm{J}_{\mathrm{exc}}$. Carrying out these steps results in the drift-diffusion equation
\begin{equation}
    \partial_t n(\bm{r},t)=\nabla\cdot (D(n) \nabla n(\bm{r},t))+\nabla\cdot (\mu_m g_{IX} n(\bm{r},t) \nabla n(\bm{r},t))-\frac{n(\bm{r},t)}{\tau} \ , 
    \label{dde}
\end{equation}
introducing the diffusion coefficient $D(n)=-\frac{1}{2A}\sum_{\bm{Q}}\tau_{\bm{Q}}v_{\bm{Q}}^2\frac{\partial N_{\bm{Q}, \mathrm{eq}}}{\partial E_{\bm{Q}}}\frac{\partial \mu}{\partial n}$ \cite{hess1996maxwell} dependent on the relaxation time $\tau_{\bm{Q}}$ and the equilibrium Bose distribution $N_{\bm{Q}, \mathrm{eq}}$ including the chemical potential $\mu=k_B T\mathrm{ln}(1-\mathrm{exp}(-T_d/T))$ with $T_d=\frac{2\pi\hbar^2 n}{k_B M}$ being the degeneracy temperature. We also defined the exciton mobility $\mu_m=-\frac{1}{2A}\sum_{\bm{Q}}\tau_{\bm{Q}}v^2_{\bm{Q}}\frac{\partial N_{\bm{Q}, \mathrm{eq}}}{\partial E_{\bm{Q}}}$ (setting $e\equiv 1$) \cite{hess1996maxwell}. In this work, we assume $\tau_{\bm{Q}}\equiv \tau$ and take $\tau$ as density-independent. In this way, we can extract the scattering time from the experimentally obtained low-density diffusion coefficient $D_0$ via the classical relation $D_0\approx{\frac{k_B T \tau}{M}}$ \cite{sun2021excitonic}. The diffusion coefficient and exciton mobility can then be analytically evaluated as $D=D_0 \frac{T_d}{T}[\mathrm{exp}(T_d/T)-1]^{-1}$and $\mu_m=\frac{D_0}{k_B T}$, respectively. Note that the diffusion coefficient $D\rightarrow D_0$ in the limit of low densities and high temperatures. A full microscopic calculation of the scattering rate $\tau$ including contributions from interlayer exciton-phonon as well as exciton-exciton scattering is beyond the scope of this work. 
When including exciton screening in the drift-diffusion equation, the exciton-exciton interaction becomes density- and spatially dependent, i.e. $g_{IX}\rightarrow g_{IX}(n)$ and $g_{IX} n\rightarrow g_{IX}(n) n+\Sigma_{CH}$, $\Sigma_{CH}=\sum_{\bm{q}}\bigg(\frac{V^{vv}_{\bm{q}}}{\epsilon_{intra, exc}(\bm{q})}-V^{vv}_{\bm{q}}\bigg)$ being the Coulomb-hole term. We also phenomenologically add a term $\propto 1/\tau$ in Eq. \eqref{dde}, reflecting the decay of interlayer excitons with the exciton life time $\tau$ extracted from a recent PL study \cite{sun2021excitonic}. All parameters used to evaluate the drift-diffusion equation are summarized in Table II in Section IV. 

\section{Input parameters}
 Table I includes the parameters used to evaluate the exciton-exciton interaction matrix elements and the density-dependent line shifts discussed in the main text. The electron and hole masses and dielectric constants are used as input parameters for the Wannier equation, providing access to the excitonic wave functions and binding energies \cite{ovesen2019interlayer}. Note that in the main text we  evaluate the interlayer exciton-exciton interaction for MoSe$_2$-WSe$_2$ heterostructures for the energetically lowest state with the electron residing in Mo-layer and the hole  in the W-layer. We also evaluate the intralayer exciton-exciton interaction in monolayer WSe$_2$. However, for completeness, we provide parameters for both WSe$_2$ and MoSe$_2$. In Table II, we provide the parameters used for simulating the interlayer exciton drift and diffusion in MoSe$_2$-hBN-WSe$_2$. 
\begin{table}[h!]
\begin{tabular}{c||c|c|c|c}

Parameter                                    & Symbol             & MoSe$_2$    & WSe$_2$ & Ref. \\ \hline
Effective electron mass (K-point)                  & $m_e$ ($m_0$)      & 0.5      & 0.29    &     \cite{kormanyos2015k} \\
Effective hole mass (K-point)                      & $m_h$ ($m_0$)      & -0.64    & -0.36   & \cite{kormanyos2015k}\\ 
Layer thickness                              & $d$ (nm)           & 0.645      & 0.672    & \cite{rasmussen2015computational} \\ 
Out-of-plane component of dielectric tensor & $\epsilon_{\perp}$ & 7.4      & 7.5     &      \cite{laturia2018dielectric} \\ 
In-plane component of dielectric tensor      & $\epsilon_{||}$    & 16.5     & 15.1    &   \cite{laturia2018dielectric}   \\ 
Lattice constant & $a_0$ (nm) & 0.33 & 0.33 & \cite{PhysRevLett.108.196802} \\ 
Effective hopping integral  & $t$ (eV) & 0.94  & 1.19 & \cite{PhysRevLett.108.196802} \\  
Single-particle band gap  & $E_g$ (eV) & 2.4 & 2.4 & \cite{kormanyos2015k} \\ \hline 
\end{tabular}
\caption{DFT input parameters used for the evaluation of exciton-exciton interaction matrix elements. The effective electron and hole masses are expressed in terms of the free electron mass $m_0$. }
\end{table}
\begin{table}[h!]
\begin{tabular}{c||c|c|c}
Parameter                   & Symbol                           & MoSe$_2$-hBN-WSe$_2$ & Ref. \\ \hline
Diffusion coefficient        & $D_0$ ($\mathrm{cm}^2/\mathrm{s}$) & 0.15                 &   \cite{sun2021excitonic}   \\ 
Temperature                  & $T$ (K)                          & 4.6                  &    \cite{sun2021excitonic}  \\
Interlayer exciton life time & $\tau$ (ns)                      & 3.5                  &   \cite{sun2021excitonic}   \\ 
Layer thickness hBN & $d_{\mathrm{hBN}}$ (nm)                      & 0.3                  &  \cite{sun2021excitonic}    \\ \hline
\end{tabular}
\caption{Parameters adapted from experimental measurements \cite{sun2021excitonic} and used for simulating interlayer exciton diffusion in MoSe$_2$-hBN-WSe$_2$.}
\end{table}
\newpage